\documentclass{natureDW} 

\newif\ifSinglePageVersion
\SinglePageVersionfalse 
\usepackage{amsmath}

\usepackage{float} 
\usepackage{verbatim}
\usepackage{graphicx,color,tikz,xspace}  
\usepackage{xr} 
\usepackage[font={footnotesize}]{caption}

\usepackage{microtype}

\usepackage[T1]{fontenc}
\usepackage[utf8]{inputenc}
\usepackage{sansmathfonts}

\usepackage{helvet}

\usepackage{hyperref}
\hypersetup{colorlinks=true,citecolor=blue,linkcolor=blue}

\makeatletter 
\def\XR@[#1]#2{{%
  \makeatletter
  \def\XR@prefix{#1}%
  \XR@next"#2.aux"\relax\\}} 
\makeatother

\makeatletter
\newcommand{\customlabel}[3]{%
   \protected@write \@auxout {}{\string \newlabel {#1}{{#2}{\thepage}{#2}{#1}{#3}} }%
}
\makeatother

\usepackage{marginnote}
\usepackage{xcolor}
\definecolor{vlightgray}{gray}{0.98}
\usepackage[textsize=small,color=vlightgray]{todonotes} 
\usepackage{textcomp}

\BiblatexSplitbibDefernumbersWarningOff

\usepackage[normalem]{ulem} 

\newcommand{\gse}{\ensuremath{G_{\mathrm{se}}}}

\newcommand{\degC}{\ensuremath{^\circ\text{C}}\xspace}
\newcommand{\muM}{\ensuremath{\text{\textmu M}}\xspace}
\newcommand{\mum}{\ensuremath{\text{\textmu m}}} 
\newcommand{\muL}{\ensuremath{\text{\textmu L}}\xspace}

\title{Pattern recognition in the nucleation kinetics of non-equilibrium self-assembly}
\author{Constantine Glen Evans$^{1,2,3}$, Jackson O'Brien$^{4}$, Erik Winfree$^1$,  Arvind Murugan$^4$ }

\hypersetup{pdftitle={Pattern recognition in the nucleation kinetics of non-equilibrium self-assembly},
            pdfauthor={Constantine Glen Evans, Jackson O'Brien, Erik Winfree, Arvind Murugan}}

\makeatletter
\AtEveryCitekey{%
  \ifcsundef{blx@entry@refsegment@\the\c@refsection @\thefield{entrykey}}
    {\csnumgdef{blx@entry@refsegment@\the\c@refsection @\thefield{entrykey}}{\the\c@refsegment}}
    {}}
\defbibcheck{onlynew}{%
  \ifnumless{0\csuse{blx@entry@refsegment@\the\c@refsection @\thefield{entrykey}}}{\the\c@refsegment}
    {\skipentry}
    {}}
\makeatother

\AtBeginBibliography{\small}

\usepackage{xparse}
\usepackage{xifthen}
\NewDocumentCommand{\figref}{mO{}}{Fig.~\ref{#1}%
\ifthenelse{\isempty{#2}}{}{#2}\xspace}

\NewDocumentCommand{\efigref}{mO{}}{Extended Data Fig.~\ref{#1}%
\ifthenelse{\isempty{#2}}{}{#2}\xspace}

\def\dataurl{https://www.dna.caltech.edu/SupplementaryMaterial/MultifariousSST/}

\def\sineural{\href{\dataurl}{Supplementary Information and Data Appendix, Section 1}\xspace} 
\def\sireservoir{\href{\dataurl}{Supplementary Information and Data Appendix, Section 1.5}\xspace} 
\def\siktam{\href{\dataurl}{Supplementary Information and Data Appendix, Section 2.1}\xspace}         
\def\sisgm{\href{\dataurl}{Supplementary Information and Data Appendix, Section 2.2}\xspace}       
\def\siwtacrn{\href{\dataurl}{Supplementary Information and Data Appendix, Section 2.3}\xspace}   
\def\simaptraining{\href{\dataurl}{Supplementary Information and Data Appendix, Section 2.4}\xspace}       
\def\siwindow{\href{\dataurl}{Supplementary Information and Data Appendix, Section 2.5}\xspace}       
\def\siimages{\href{\dataurl}{Supplementary Information and Data Appendix, Section 2.7}\xspace}       
\def\sialltraining{\href{\dataurl}{Supplementary Information and Data Appendix, Sections 2.4 and 2.5}\xspace}       
\def\sipatmix{\href{\dataurl}{Supplementary Information and Data Appendix, Section 2.8}\xspace}   
\def\sifluorophores{\href{\dataurl}{Supplementary Information and Data Appendix, Section 3}\xspace}    
\def\sitiles{\href{\dataurl}{Supplementary Information and Data Appendix, Sections 3.3 and 4.1}\xspace}  
\def\sioxDNA{\href{\dataurl}{Supplementary Information and Data Appendix, Section 3.4}\xspace} 
\def\siflagdata{\href{\dataurl}{Supplementary Information and Data Appendix, Section 5}\xspace}   
\def\siAflag1{\href{\dataurl}{Supplementary Information and Data Appendix, Section 5.3.13}\xspace}   
\def\sipatterndata{\href{\dataurl}{Supplementary Information and Data Appendix, Section 6}\xspace}       
\def\sicountdata{\href{\dataurl}{Supplementary Information and Data Appendix, Section 6.3}\xspace}   
\def\simockingbird{\href{\dataurl}{Supplementary Information and Data Appendix, Section 6.4.9}\xspace}    
\def\sitprots{\href{\dataurl}{Supplementary Information and Data Appendix, Sections 5 and 6}\xspace}  
\def\sisimplertraining{\href{\dataurl}{Supplementary Information and Data Appendix, Section 2.6}}

\begin{document}
\begin{refsegment}

\twocolumn[{%
\maketitle

\vspace{\baselineskip}{{\em \noindent 
${}^1$California Institute of Technology, Pasadena, CA,
${}^2$Evans Foundation for Molecular Medicine, Pasadena, CA, 
${}^3$Maynooth University, Maynooth, Ireland,
${}^4$University of Chicago, Chicago, IL}}
\vspace{\baselineskip}


\begin{abstract}
   {\bf
Inspired by biology's most sophisticated computer, the brain, neural networks constitute a profound reformulation of computational principles\autocite{HKP1991,dayan2005theoretical,goodfellow2016deep}. Remarkably, analogous high-dimensional, highly-interconnected computational architectures also arise within information-processing molecular systems inside living cells, such as signal transduction cascades and genetic regulatory networks\autocite{rossler1974synthetic,Hjelmfelt1991-uq,mjolsness1991connectionist,Bray1995-yf}. Might collective modes analogous to neural computation be found more broadly in other physical and chemical processes, even those that ostensibly play non-information-processing roles?
Here we examine nucleation during self-assembly of multicomponent structures, showing that high-dimensional patterns of concentrations can be discriminated and classified in a manner similar to neural network computation. Specifically, we design a set of 917 DNA tiles that can self-assemble in three alternative ways such that competitive nucleation depends sensitively on the extent of colocalization of high-concentration tiles within the three structures. The system was trained in-silico to classify a set of 18 grayscale 30×30 pixel images into three categories. Experimentally, fluorescence and atomic force microscopy measurements during and after a 150-hour anneal established that all trained images were correctly classified, while a test set of image variations probed the robustness of the results. While slow compared to prior biochemical neural networks, our approach is surprisingly compact, robust, and scalable. Our findings suggest that ubiquitous physical phenomena, such as nucleation, may hold powerful information processing capabilities when they occur within high-dimensional multicomponent systems. 
   }
\end{abstract}
\vspace{4ex}

\vspace{0.2cm}

}]

\begin{figure}[h!!!] 
	\centerline{\includegraphics[width=89mm]{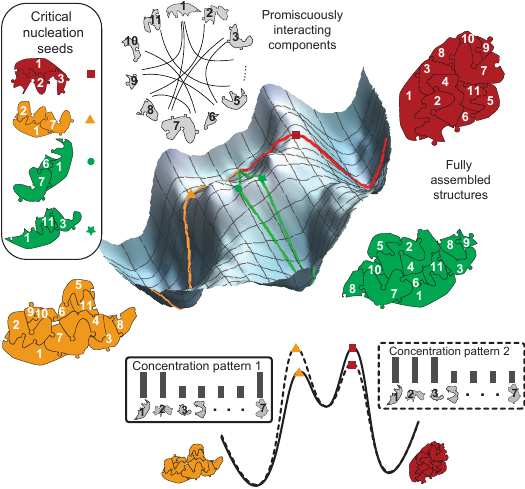}}
	\caption[]{
	\textbf{Conceptual framework for pattern recognition by nucleation.} When one set of molecules can potentially assemble multiple distinct structures, the nucleation process that selects between outcomes is responsive to high-dimensional concentration patterns. 	Assembly pathways can be depicted on an energy landscape (schematic shown) as paths from	a basin for unassembled components that proceed through critical nucleation seeds (barriers) to a basin for each possible final structure. Seeds that colocalize high concentration components will lower the nucleation barrier for corresponding assembly pathways. The resulting selectivity of nucleation in high-dimensional self-assembly is sufficiently expressive to perform complex pattern recognition in a manner analogous to neural computation
	(see \efigref{extfig:neural}).
} \label{fig:concepts}
\end{figure}

The success of life on earth derives from its use of molecules to carry information, implement algorithms that control chemistry and respond intelligently to the environment.   Genetic information encodes not only molecules with structural and chemical functionality, but also biochemical circuits that in turn process internal and external information relevant for cellular decision-making.  
While some biological systems may, like modern modular engineering, isolate information processing from the physical subsystems being controlled\autocite{hartwell1999modular}, other critical decision-making may be embedded within and inseparable from processes such as protein synthesis, metabolism, self-assembly, and structural reconfiguration. 
Understanding such physically-entangled computation is necessary not only for understanding biology, but also for engineering autonomous molecular systems such as artificial cells, where it is essential to pack as much capability as possible within limited space and energy budgets.

The interplay of structure and computation is particularly rich in molecular self-assembly.  In biological cells, decisions about navigation, chemotaxis, and phagocytosis are made via structural rearrangements of the cytoskeleton that integrate mechanical forces and chemical signals\autocite{fletcherCellMechanicsCytoskeleton2010,Holy1994-hm,lee2015coccidioides,floyd2021cytoskeletal_avalanches}, but where and how information processing occurs remains elusive.  
In DNA nanotechnology\autocite{seemanDNANanotechnology2018}, self-assembly of DNA tiles has been shown theoretically and experimentally to be capable of Turing-universal computation through simulation of cellular automata and Boolean circuits\autocite{Rothemund2000squares,rothemund2004algorithmic,woods_doty_diverse_2019}, but this digital model of computation lacks a clear analog in biology. 

\begin{figure*}[t!]
	\centerline{\includegraphics[width=183mm]{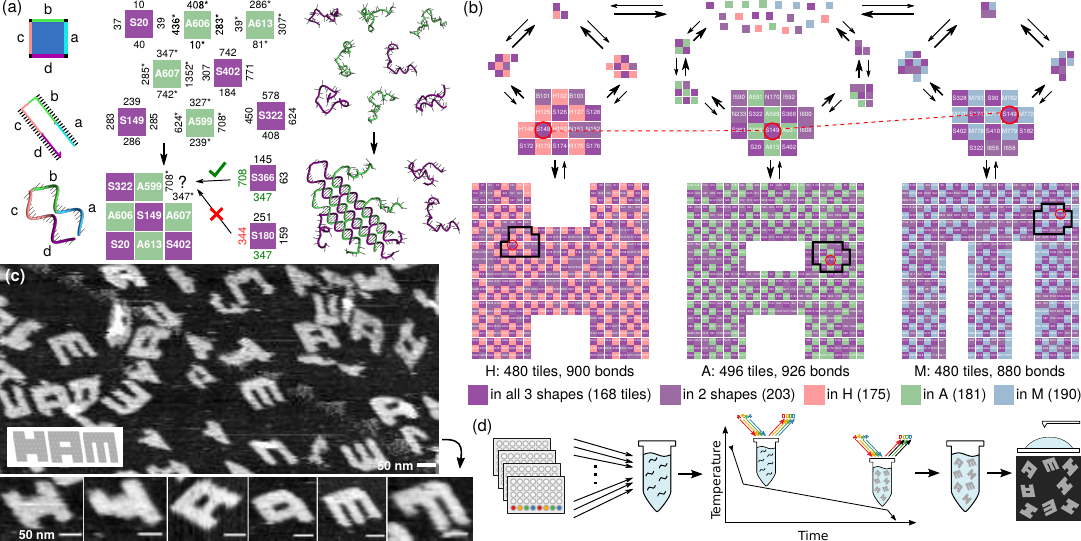}}
	\caption[]{
      \textbf{A multifarious mixture of 917 molecular species that can assemble into three distinct structures from one set of molecules.} \textbf{a,} 42-nucleotide DNA strands self-assemble into two-dimensional (2D) structures by forming bonds with four complementary strands in solution via four 10 or 11 nucleotide domains. The strands can be abstracted as square tiles, each named and shown with distinct binding domains identified by number, such that e.g. $708$ is complementary to $708^*$. At nucleation and growth temperatures, attaching by two bonds or more is favorable, while one is insufficient. \textbf{b,} One pool of 917 tile types assembles into three distinct shapes, H, A and M, through a multitude of pathways.  While each tile occurs at most once in each shape, the shared purple species recur in multiple shapes, in distinct spatial arrangements; e.g., S149 is highlighted in red. \textbf{c,} Annealing an equal mix of all tiles results in a mixture of fully and partially assembled H, A and M, imaged by atomic force microscope (AFM).  This is the same sample as ``SHAM60'' in \figref{fig:patternexp}[e]. Inset illustrates the expected slant of the shapes due to single-stranded tile geometry.
      \textbf{d,} A typical experiment mixes some concentration of each tile type into a single tube, with some tiles swapped for fluorophore- and quencher-modified versions.  The sample is heated to remove any preexisting binding, cooled to a temperature slightly above where any growth is observed, then slowly annealed through a small range of temperatures while fluorescence is measured in a qPCR machine; samples are then imaged by AFM.
	} \label{fig:multifarious}
\end{figure*}

Neural computation is an alternative form of naturally compact computation with several distinctive hallmarks~\autocite{HKP1991,dayan2005theoretical,goodfellow2016deep}:  mixed analog/digital decision-making, recognition of high-dimensional patterns, reliance on the collective influence of many distributed weak interactions, robustness to noise, and an inherent ability to learn and generalize.  
A paradigmatic neural network model is the Hopfield associative memory\autocite{Hopfield1982-yy}, which conceptualizes dynamics as a random walk on an energy landscape that has been sculpted by learning to contain attractor basins at each memory.
Surprisingly, neural network models map naturally onto models of well-mixed chemical networks\autocite{rossler1974synthetic,Hjelmfelt1991-uq}, genetic regulatory networks~\autocite{mjolsness1991connectionist}, and signal transduction cascades\autocite{Bray1995-yf}; 
such networks have been experimentally demonstrated both in cell-free systems and within living cells~\autocite{Qian2011-jh,Cherry2018-qf,okumura2022nonlinear,rizik2022synthetic}.  But these well-mixed approaches still separate decision making from downstream processes.

Neural information processing principles embedded within molecular self-assembly have been harder to discern, and perhaps at first appear as a contradiction in terms.  An early thermodynamic view of how free-energy minimization in molecular self-assembly could be akin to the Hopfield model did not lead to concrete realizations~\autocite{conradSelfassemblyMechanismMolecular1989}.  However, a recent kinetic view of multicomponent systems that permit assembly of many distinct structures using the same components (“multifarious self-assembly”)~\autocite{Murugan2015-ps,zhong_associative_2017} revealed concrete connections to Hopfield associative memories~\autocite{Hopfield1982-yy}  and models of hippocampal place cells~\autocite{moserPlaceCellsGrid2008} at the level of collective dynamics, even though individual molecules do not explicitly mimic the mechanistic behavior of individual neurons. 

Here, we reformulate this connection as an intrinsic feature of heterogeneous nucleation kinetics and experimentally demonstrate its power for high-dimensional pattern recognition using DNA nanotechnology\autocite{seemanDNANanotechnology2018}.
The phenomenon arises when the same components can form several distinct assemblies in different geometric arrangements (\figref{fig:concepts}).  Nucleation proceeds by spontaneous formation of a critical seed that subsequently grows into a structure \autocite{frenkel}. Because the nucleation rate of a seed depends strongly on the bulk concentrations of components that occur in that seed, and many distinct seeds and pathways may be viable, the overall rate of formation of a given structure is a complex function of the concentration pattern.  Further, because components are shared between structures, competition for resources~\autocite{genotComputingCompetitionBiochemical2012a} results in a winner-take-all effect that accentuates the discrimination between concentration patterns.

\vspace{0.2cm}  \noindent {\bf \large Molecular system design}  \vspace{0.1cm}

To explore these principles experimentally, we take advantage of the powerful foundation provided by DNA nanotechnology for programming molecular self-assembly.  The well-understood kinetics and thermodynamics of Watson-Crick base pairing enables systematic sequence design
~\autocite{seemanNovoDesignSequences1990,zadeh2011nupack} for DNA tiles that reliably self-assemble into 
periodic, uniquely-addressed, and algorithmically-patterned structures with 100s to 1000s of distinct tile types\autocite{Winfree98nature,rothemund2004algorithmic,yin_programming_2008,wei_complex_2012,ke_three-dimensional_2012,ong2017programmable,woods_doty_diverse_2019}. 
These classes of self-assembly differ in the structures produced and in the nature of interactions: in periodic and uniquely-addressed structures, each molecular component typically has a unique possible binding partner in each direction. For algorithmic patterns (as for multifarious assembly), some components have multiple possible binding partners, such that which one attaches at a given location is decided during self-assembly based on which forms more bonds with neighboring tiles.

We build on these ideas to create a molecular system capable of assembling multiple target structures (H, A, and M in \figref{fig:multifarious})
from a shared set of interacting components by colocalizing them in different ways. The first stage of design begins with a set S of shared tiles that do not directly bind each other; then three sets of interaction mediating-tiles (also called H, A, and M) are introduced for each of the respective desired structures. Each interaction tile in, e.g. H, binds four specific S tiles together in a checkerboard arrangement that reflects neighborhood constraints between shared S tiles in structure H. 
These H interaction tiles are unique to structure H and do not occur in the assembled A or M structures.

Tiles in a 1:1 stoichiometric mix of S+H, S+A or S+M will have no promiscuous interactions and will assemble H, A or M respectively, as with prior work on uniquely addressable structures~\autocite{wei_complex_2012}. But a 1:1:1:1 mix of S+H+A+M, henceforth called our SHAM mix, is capable of assembling three distinct structures. This additive construction of interaction-mediating tiles is analogous to Hebbian learning of multiple memories in Hopfield neural networks~\autocite{Hopfield1982-yy,Murugan2015-ps}; see \efigref{extfig:neural}. 
Furthermore, the use of interaction-mediating tiles avoids constraints from Watson-Crick complementarity, allowing almost arbitrary interactions to be engineered between S tiles. To avoid undesired consequences of the extensive promiscuous interactions present in the SHAM mix, later design stages optimized this initial layout using self-assembly proofreading principles to reduce errors\autocite{winfree_proofreading_2004,evans_optimizing_2018}; see Extended Data Figs. \ref{extfig:merging} and \ref{extfig:errorsims}. 

The resulting design in \figref{fig:multifarious}[b], has 168 tiles shared across all three shapes, 203 tiles shared across a pair, and 546 tiles unique to a specific shape. Our experimental implementation used 42-nucleotide single-stranded DNA tiles\autocite{wei_complex_2012} (\figref{fig:multifarious}[a]) with sequences designed using tools from prior work\autocite{woods_doty_diverse_2019} to reduce unintended interactions and secondary structure and ensure nearly-uniform binding energies. 

To test whether proofreading was sufficient to combat promiscuity and to test the unbiased yield of different structures, we annealed all tiles at equal concentration (60 nM) in solution over 150 hours from 48\,\degC to 45\,\degC.  
Atomic force microscopy (AFM) revealed a roughly equal yield of all three structures (\figref{fig:multifarious}[c]).
Despite being a slow anneal, this uniform distribution is incompatible with an equilibrium Boltzmann distribution that would exponentially magnify differences in the area and perimeter (and thus free energy) of H, A and M; but it is compatible with kinetically controlled assembly where nucleation rates are linearly proportional to a shape's area, as nucleation could occur anywhere within the shape.
Additionally, we did not observe significant chimeric structures or uncontrolled aggregation, indicating that proofreading was functioning as desired.  However, many structures appeared to be incomplete, often missing tiles from two specific corners, perhaps due to asymmetric growth kinetics or lattice curvature\autocite{yin_programming_2008},
or (in the case of A only) exhibited signs of spiral defect growth (\efigref{extfig:errorsims}).

\begin{figure*}[t!!!]
	\centerline{\includegraphics[width=183mm]{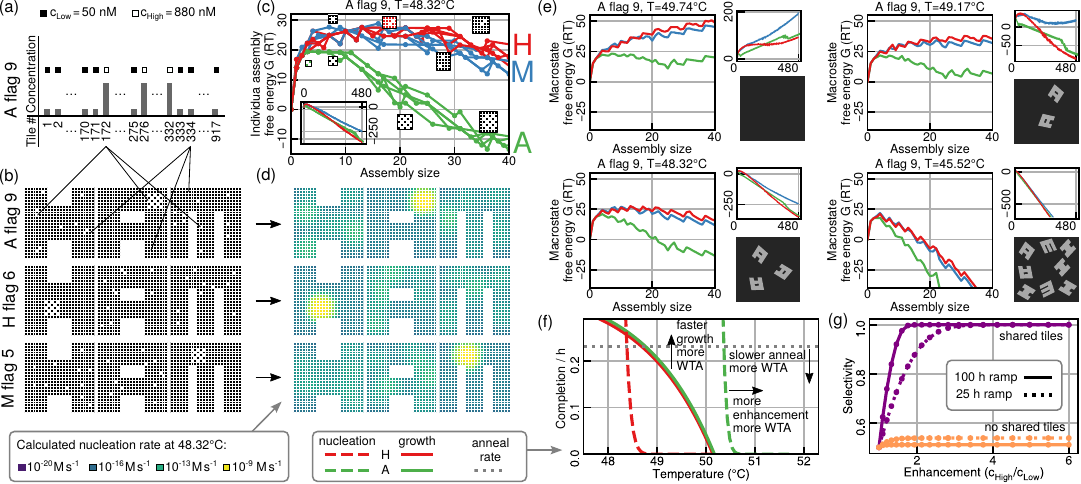}}
	\caption[]{\textbf{Theory shows selective nucleation when high concentration tiles are co-localized in one shape more than in others.}
	\textbf{a,} One pattern (``A flag 9'') enhancing the concentration of shared tiles colocalized in A but relatively dispersed in H and M. \textbf{b,} A flag 9 plotted by tile locations in each shape, along with example ``flag'' patterns that have colocalization in H and M. \textbf{c,} For A flag 9, free energies of assemblies along predicted nucleation pathways for each shape (\efigref{extfig:nucleation}). 
    Several example assemblies are shown; the green and red ones are critical seeds for the A and H pathways respectively.
	\textbf{d,} Regions predicted to participate in nucleation by the simulation for three concentration patterns (lighter colors correspond to higher participation). \textbf{e,} Macrostate free energies for sets of partial assemblies of increasing size (number of tiles) and predicted AFM results at several temperatures spanning the melting temperature. 
     Small plots show the full size range, thus
    illustrating the independence of the nucleation barrier kinetics and the complete assembly thermodynamics.
	\textbf{f,} 
    For on-target (A, green) and off-target (H, red) shapes, nucleation rates (dashed) and growth rates (solid) are 
    plotted as a function of temperature, according to the simplified model of \efigref{extfig:nucleation}[f].  Rates are given relative to the time to completely consume the lowest-concentration tile; the horizontal dotted line indicates the rate of annealing between the on-target to off-target nucleation temperatures.
    Due to the higher nucleation temperature for the on-target shape, when annealing time scales are comparable to or slower than growth time scales, depletion of shared tiles during a temperature anneal can lead to a winner-take-all (WTA) effect.  Slower annealing and faster growth can increase the WTA effect.    \textbf{g,} In this model, WTA leads to higher selectivity (on-target vs. total nucleation) compared to systems with no shared components; for slower anneals, selectivity increases for systems with shared components, but decreases for systems with no shared components.
 } \label{fig:nucleation}
\end{figure*}

\vspace{0.2cm} \noindent {\bf \large Colocalization controls nucleation} \vspace{0.1cm} 

Understanding nucleation in multicomponent self-assembly has required extensions of classical nucleation theory\autocite{frenkel} that have effectively guided the design of programmable DNA tile systems with well-defined assembly pathways\autocite{schulman_programmable_2009,schulman2007synthesis,jacobs2016self,sajfutdinow2018nucleation}.
Building on this work, here we examine how selection between target structures that differ in colocalization of tiles can be determined by nucleation kinetics and controlled by concentration patterns.
We model the free energy of a structure $A$ with $B$ total bonds {as} $G(A) = \sum_{i \in A} G_{mc}^i - B\,G_{se} -
\alpha$, where $\alpha$ depends on the choice of reference concentration $u_0$, $G_{mc}^i = \alpha -\log c_i/{u_0}$ is the chemical potential (or equivalently, translational entropy) of tile $i$ at concentration $c_i$, and $G_{se}$ is the energy of each bond in units of $R T$. 
$G(A)$ has competing contributions that scale with the structure's area and perimeter, and is hence maximized for certain partial assemblies called critical nucleation seeds.
The formation of such seeds is often rate-limiting: once these seeds are assembled, subsequent growth is faster and mostly `downhill' in free energy. 
If the nucleation rate $\eta_{shape}$ for a given shape is dominated by a single critical nucleus $A_s$, we could use an Arrhenius-like approximation $\eta_{shape} \sim e^{-G(A_{s})}$; in the case that multiple critical nuclei are significant, we must perform a sum.

When such analyses are applied to homogeneous crystals with uniform concentration $c_i = c$ of components, critical nuclei are simply those with the appropriate balance of size and perimeter. Heterogeneous concentration patterns require a more nuanced analysis: critical seeds can now be arbitrarily shaped, potentially offsetting a larger perimeter penalty by incorporating tiles with higher bulk concentration.
Therefore, we implemented a stochastic sampling algorithm to estimate the nucleation rate of a structure with an uneven pattern of concentrations (\efigref{extfig:nucleation}). 

Consider the examples in \figref{fig:nucleation} where the concentrations of some shared tiles in the SHAM mix have been enhanced. These high concentration tiles are colocalized in structure A but scattered across H and M. Consequently, such a pattern will lower kinetic barriers for the nucleation of A while maintaining high barriers for H and M. The typical area $K$ over which colocalization promotes nucleation can be estimated from the size of critical seeds predicted by classical nucleation theory and is generally larger at higher temperatures\autocite{frenkel}. Hence we expect a trade-off between speed and complexity of pattern recognition (\figref{fig:nucleation}[e]), with more subtle discrimination at higher temperatures (large $K$)---at the expense of slower experiments---and lower discriminatory power at lower temperatures (small $K$).

\begin{figure*}[t!!!]
	\centerline{\includegraphics[width=183mm]{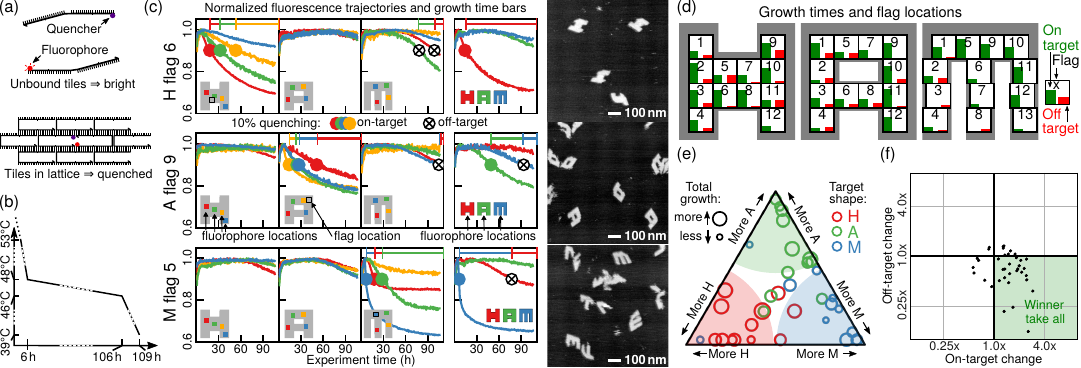}}
	\caption[]{
      {\bf Selective nucleation in experiments with shape-specific localized concentration patterns of shared tiles.}
      {\bf a,} Pairs of alternative tiles with a fluorophore and quencher (\figref{fig:multifarious}[d]) have their fluorescence quenched when incorporated together in an assembly; small assemblies of just a few strands do not effectively quench
      (see \efigref{extfig:fluorophores}).
      {\bf b,} Samples were annealed with a temperature protocol that 
      cooled from 71\,\degC (well above melting temperature) to 48\,\degC over $\sim 6$~hours, 
      cooled to 46\,\degC over 100~hours, and finally cooled to 39.5\,\degC over 3~hours (see \efigref{extfig:flags}).
     {\bf c,} Experimental results for the 3 flag patterns shown in \figref{fig:nucleation}.  The positions of fluorophore/quencher tile pairs used in each of the four samples are shown by the inset icons. Points where fluorescence signals dropped by 10\% below their maximum (to which signals were normalized) are shown with colored dots for on-target nucleation and with $\otimes$ for off-target nucleation. `Growth times' measure the period from `10\% quenching' to the end of the experiment, shown as horizontal bars. Sample AFM images from one of the samples are shown for each flag.
      {\bf d,} Total growth times for on-target versus off-target nucleation are summarized for all 37 flag patterns.
      Each numbered box indicates the location of the corresponding $5 \times 5$ checkerboard flag; good performance is indicated by a tall green bar and a short red bar.
      {\bf e,} The same data displayed as a ternary plot, with proximity to triangle corners indicating relative fractions of growth time and circle size indicating overall growth time.
      {\bf f,} Average change in quenching (a measure of nucleation) of on- and off-target structures with flag patterns compared to equimolar SHAM mixes. 
      Each dot represents a single flag pattern (see \efigref{extfig:wta}).
      For most patterns, increasing shared tile concentrations reduces the absolute off-target nucleation, supporting a winner-take-all effect.
      \label{fig:flagscan}}
\end{figure*}

To experimentally characterize the basis of selectivity, we systematically tested a series of 37 concentration patterns, which we call `flags' because each one uses high concentrations in a checkerboard localized somewhere in one of the shapes (three examples are shown in \figref{fig:nucleation}[b]). We did not enhance concentrations of tiles unique to shapes, to avoid additional thermodynamic bias towards any one structure.
We ramped the temperature down slowly, from 48\,\degC to 46\,\degC (the expected range 
for nucleation, a few degrees below the melting temperatures) to provide robustness to variations in nucleation temperatures among flags in different locations and to probe for slow off-target nucleation.
To monitor nucleation and growth in real time, we designed distinct fluorophore/quencher pairs on adjacent tiles in four locations on each shape, using tiles not shared between shapes.  Each pair quenches when the local region of that specific structure assembles (\figref{fig:flagscan}[a]).

Experimental results illustrating selective nucleation are shown in \figref{fig:flagscan}[c] for three example flag concentration patterns.  When the pattern localizes high concentration species in a structure, e.g., H, the fluorophore in the expected nucleation region of that structure quenched first and rapidly. 
After a delay, fluorophore signals from other parts of the same structure also dropped, indicating growth. Fluorophores on off-target structures show minimal to no quenching until late in the experiment.  AFM images from samples at the end of the experiment confirm that fluorophore quenching corresponded to selective self-assembly of complete or partial shapes.  Of the 37 flag positions, roughly half exhibited robust selective nucleation and growth (\figref{fig:flagscan}[de]), while other positions were either not selective or did not grow well, for reasons we have not been able to determine.

In multifarious systems, we expect enhanced selectivity because of a competitive suppression of nucleation.  Using an annealing protocol that spends sufficient time at temperatures where A can nucleate and grow significantly, but H cannot nucleate (\figref{fig:nucleation}[f]), we expect a winner-take-all (WTA) effect in which the assembly of A depletes shared tiles S and thus actively suppresses nucleation of H. As shown in \figref{fig:flagscan}[f], we see evidence for such a WTA effect in most experiments. 
Such a winner-take-all effect can enhance the effect of small differences in nucleation kinetics.

\vspace{0.2cm} \noindent {\bf \large Pattern recognition by nucleation}  \vspace{0.1cm} 

\begin{figure*}[t!!!]
	\centering
	\includegraphics[width=183mm]{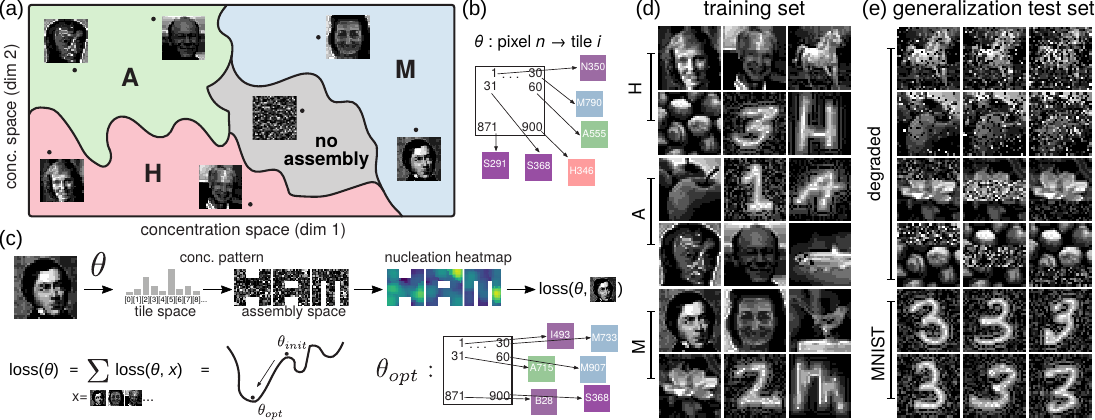}
	\caption[]{
     \textbf{Design of self-assembly phase diagrams to solve pattern recognition problems.} \textbf{a,} Phase diagram shows desired outcomes of kinetically controlled self-assembly in different regions of $N=917$ dimensional concentration space (2D schematic shown).  Each grayscale image represents a vector of tile concentrations. \textbf{b,} $\theta$ specifies which pixel location corresponds to which tile. \textbf{c,} Given a map $\theta$, any image can be converted to a tile concentration vector by associating the grayscale value of pixel location $n$ with the concentration of the corresponding tile $i=\theta(n)$. We compute the `loss' for a given pixel-to-tile map $\theta$ using simulations to estimate the nucleation rates of desired and undesired structures for each image and summing over a training set. Stochastic optimization in $\theta$ space gives a putative optimal $\theta_{opt}$ that we used for experiments. \textbf{d,} Images used for training. \textbf{e,} Additional images used to test generalization power. 
     \label{fig:phasedesign}}
\end{figure*}

Our work thus far shows that the space of all concentration patterns $\mathbf{C} = \mathcal{R}^N$, which includes patterns not experimentally tested, consists of regions that result in the selective assembly of each of H, A and M respectively (\figref{fig:phasedesign}[a]).  These regions together represent a phase diagram for this self-assembling system~\autocite{Murugan2015-ps} that reflects the decisions it makes to classify concentration patterns.  While phase boundaries of traditionally studied physical systems are usually low dimensional and not fruitfully interpreted as decision boundaries, in multicomponent heterogeneous systems like ours, the phase diagram is naturally high dimensional. More generally, phase boundaries in disordered many-body systems tend to be complex and thus implicitly solve complex pattern recognition problems, a perspective that also underlies Hopfield's associative memory in neural networks\autocite{Hopfield1982-yy,Amit1985-as}. 

Here, nucleation is solving a particular pattern recognition problem based on which molecules are colocalized in different structures. 
Similar colocalization-based decision boundaries arise in neural place cells studied by the Mosers\autocite{moserPlaceCellsGrid2008, Battaglia1998-kd,zhong_associative_2017,Monasson2015-nl} and are 
complex enough to solve pattern recognition problems and permit statistically robust learning.

Having demonstrated that multifarious self-assembly can solve a specific pattern recognition problem, could different molecules be designed to solve other tasks such as recognizing or classifying images?
Here, the grayscale value of each pixel position in the $30 \times 30$ images is taken to represent the concentration of a distinct molecule. Instead of synthesizing new molecules with new interactions to solve the above challenge, we show that the design problem is solvable with our existing molecules by an optimized choice of a pixel-to-tile map $\theta$ that specifies which existing tile should correspond to which pixel position (\figref{fig:phasedesign}[b]). In addition to saving DNA synthesis costs, this approach helps demonstrate that a random molecular design can be exploited, \textit{ex post facto}, to solve a specific computational problem by modifying how the problem is mapped onto physical components, as done in reservoir computing \autocite{tanaka2019reservoir}.

We specified our design problem by picking  arbitrary images as training sets shown in \figref{fig:phasedesign}[d]. Note that images in one class share no more resemblance than images across classes, e.g., class H is Hodgkin, Hopfield, Horse etc, though the number of pixels and grayscale histogram were standardized across images (see Methods). In this way, the number of distinct images per class (6 in the experiments presented below) tests the flexibility of decision surfaces inherent to this self-assembling molecular system as a classifier.

We then used an optimization algorithm (see \figref{fig:phasedesign}[c] and Methods) on $\theta$ that sought to maximize nucleation of the on-target structure for the concentration pattern corresponding to each image while also minimizing off-target nucleation.
That is, our algorithm sought to map high concentration pixels in each image (e.g., Mitscherlich) to colocalized tiles in the corresponding on-target structure (here, M) to enhance nucleation, while mapping those same pixels to scattered tiles in undesired structures (here, A and H). Note that this map $\theta$ is simultaneously optimized for all images and not independently for each image. Hence no map $\theta$ might be able to perfectly satisfy all the above requirements simultaneously for all images in all classes; analogous to associative memory capacity~\autocite{Hopfield1982-yy, Amit1985-as,Murugan2015-ps}, performance drops as one attempts to train more patterns (\efigref{extfig:capacity}).

\begin{figure*}[t!!]
    \centering
    \includegraphics{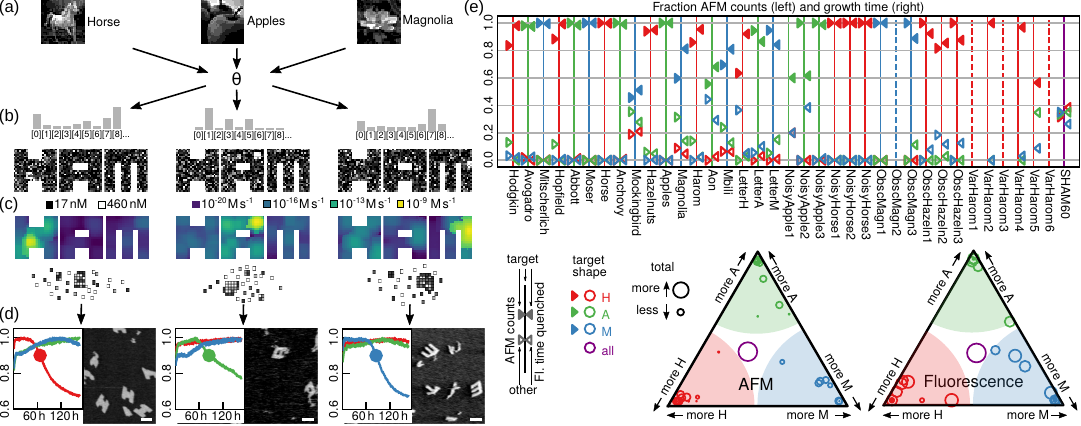}
    \caption{\textbf{Pattern recognition results with a pre-existing multifarious system.} \textbf{a--c,} All images (three shown) are converted (\textbf{a}) through a single pixel-to-tile map $\theta$ to vectors of tile concentrations (\textbf{b}), which are shown mapped onto tile locations in each shape, with nucleation rate predictions (\textbf{c}). \textbf{d,} Normalized fluorescence over time in hours (one label per shape; other label configurations shown in \efigref{extfig:patterns}) during a 150 hour temperature ramp from 48\,\degC to 45\,\degC, and final AFM images (with 100 nm scale bars).  \textbf{e,} Summary of results for both fluorescence and AFM for all 36 images, and a uniform 60 nM tile concentration control sample. Above, colors of vertical lines indicate the target shape for each pattern, while triangular markings of each color indicate the relative fraction of growth time (on right) or fraction of shapes counted in AFM images (on left) for the corresponding shape (solid markings indicate target shape).  Dashed lines indicate samples with no significant quenching or observed shapes.  Below, ternary plots summarize the same results, with proximity to triangle corners indicating relative fractions of growth time (right) or counted shapes (left), and circle size indicating overall growth time (right) or total number of shapes (left).}
    \label{fig:patternexp}
\end{figure*}

For pattern recognition experiments, we enhanced concentrations of tiles in the SHAM mix in accordance to each of the 18 training images (using the optimized $\theta$) and annealed each of the 18 mixes with a 150 h ramp from 48\,\degC to 45\,\degC. 
As verified by AFM imaging and real-time fluorescence quenching, we found that the 18 training images yielded correct nucleation, in the sense that there was more of the correct shape than any other shape, and in all but 5 cases was highly (more than 80\%) selective 
(\figref{fig:patternexp}).

We also tested 12 degraded images and 6 alternate handwriting images (\figref{fig:phasedesign}[e]), with the same trained pixel-to-tile map $\theta$. Pattern recognition was successful for random speckle distortions and all but one partly obscured image.  Generalization, the ability to recognize related images not present in a training set, is a critical aspect of learning in neural networks. A given architecture can be naturally robust to certain families of distortions (e.g., convolutional networks can handle translation) but not others (e.g., dilation). Since nucleation is a cooperative process, often dominated by one or a few critical seeds involving just a handful of tiles, flipping of random uncorrelated pixels and obscuring parts of an image that do not involve those critical pixel combinations will not inhibit nucleation -- demonstrating robustness. 
On the other hand, only 3 of the 6 alternate handwritten digits were correctly recognized by self-assembly, indicating a lack of robustness to this type of variation without further training.

\vspace{0.2cm} \noindent {\bf \large Discussion}  \vspace{0.1cm}

The phenomena underlying pattern recognition by multifarious self-assembly may be exploited by complex evolved or designed systems (\efigref{extfig:biology}). 
Beyond self-assembly, molecular folding processes could potentially recognize patterns in the concentrations of cofactors or subcomponents if folding kinetics can select between distinct stable states\autocite{dunn2015guiding}.
Similarly, the phase boundaries for multicomponent condensates governing genetic regulation\autocite{Hnisz2017-kg} may also contain inherent information-processing capabilities. 
In such cases, the `pixel-to-tile' map would instead correspond to a layer of phosphorylation or binding circuitry that activates or deactivates specific components based on the levels of upstream information-bearing molecular signals.  Within artificial cells\autocite{pirzer2022tiny}, multicomponent nucleation may be an especially compact way to implement decision-making within the limited space constraints.

To better understand the information-processing potential of nucleation, we may treat this physical process as a machine learning model. A key issue is how the complexity of decision surfaces, quantified in terms of computational power or learning capacity, depends on underlying physical aspects of self-assembly such as the number of molecular species, binding specificity, and geometry\autocite{minev2021robust,wintersinger2023multi}. Our work already suggests that temperature mediates a trade-off between speed, accuracy, and complexity of pattern recognition; at higher temperatures, nucleation seeds are larger, allowing discrimination based on higher order correlations in the concentration patterns, but the physical process is also correspondingly slower.  The trade-off derives from how computation here exploits the inherently stochastic nature of nucleation: monomers must make many unsuccessful attempts at forming a critical seed for both on- and off-target structures, with repeated disassembly before discovering the seed for the correct pattern recognition outcome. Relating such backtracking to stochastic search algorithms for NP-complete problems, as has been done for well-mixed chemistry\autocite{winfree2019crnsat}, might characterize the computational power of stochastic nucleation.

Viewing nucleation as a machine learning model raises the question of whether there is a natural physical implementation of learning.
Here, we trained decision boundaries \emph{in silico} using ideas from reservoir computing\autocite{tanaka2019reservoir, wrightDeepPhysicalNeural2022}: 
molecules with a fixed set of interactions could nevertheless solve an arbitrary problem by changing the mapping between inputs and fixed components (see \efigref{extfig:neural}). The analogy between Hopfield associative memories and multifarious self-assembly, especially those based on random colocalization\autocite{Murugan2015-ps,zhong_associative_2017, moserPlaceCellsGrid2008, Battaglia1998-kd, Monasson2015-nl}, suggests a way to go beyond fixed components to a scenario where interactions between components are learned in a Hebbian manner by a natural physical process. 
Notably, interactions between shared tiles in our system are mediated by shape-specific molecules. If these interaction-mediating tiles could be physically created or activated in response to environmental inputs, e.g., through proximity-based ligation, molecular systems could autonomously learn new self-assembling behaviors from examples\autocite{2016LakinStefanovic} without the need for computer-based learning. Alternatively, the natural evolution of hydrophobic residues to stabilize multi-protein complexes may have the necessary properties for inducing multifarious pattern recognition\autocite{hochberg2020hydrophobic}.

The connection between self-assembly and neural network computation raises many questions for further exploration, the broadest being a variant on Anderson’s observation that `more is different' \autocite{anderson1972more}.  
Anderson was referring to the fact that systems containing many copies of the same simple component can show emergent phenomena, such as fluid dynamics, that are best understood at a higher level.  Biology also explores another sense of `more is different’: it often makes use of a few copies of a great many different types of components\autocite{hartwell1999modular}.   
Here, new phenomena naturally emerge in the ‘large $N$ limit’: robustness, programmability, and information-processing.  
These phenomena are best explored in information-rich model systems devoid of the distracting complexities of biology. DNA nanotechnology provides one such platform that already hints at such  ‘more types is different’ phenomena. For example, self-assembled few-component DNA structures are often sensitive to sequence details and molecular purity, thus taking years to refine experimentally, while DNA origami\autocite{Rothemund2006-ca} and uniquely-addressed tile systems \autocite{wei_complex_2012,ke_three-dimensional_2012,ong2017programmable} use 100s to 1000s of components and usually work on the first try, even with unpurified strands, imprecise stoichiometry, and no sequence optimization.  
Such observations suggest heterogeneity as a defining principle for biological self-assembly\autocite{sartori2020lessons}. 

Our work adds sophisticated information-processing as a new emergent phenomenon in which self-assembly, in the multicomponent limit, gains programmable and potentially learnable phase boundaries to solve specific pattern recognition problems, analogous to earlier results for large $N$ neural networks~\autocite{Amit1985-as}. This neural network inspired perspective may help us recognize information processing in high-dimensional molecular systems that is deeply entangled within physical processes, whether in biology or in molecular engineering: multicomponent liquid condensates, multicomponent active matter, and other systems might have similar programmable and learnable phase boundaries.

\printbibliography[segment=1,check=onlynew]
\end{refsegment}
\newpage
\begin{refsegment}

\begin{methods}
\newcommand{\fm}[1]{{\color{red} #1}}

\subsection*{Multifarious DNA tile system design.}
Prior theoretical proposals\autocite{Murugan2015-ps,Zhou2015-ym} for multifarious mixtures require each component to accept multiple strongly binding partners at each binding site. However, in DNA tile assembly, each binding site can usually only bind its Watson-Crick complement, not an arbitrary set of other domains.
Hence, we used an alternate approach: we laid out three structures made of entirely unique, abstract tiles, designed a merging algorithm to reuse tiles in multiple locations if consequences for unintentional binding between other tiles was minimal, and then designed DNA sequences reflecting the resulting abstract layout of tiles. 

The three target shapes were drawn on a 24 $\times$ 24 single-stranded tile (SST) molecular canvas\autocite{wei_complex_2012}, at an abstract level without sequences.  Each location in each shape was initially a unique tile, with four abstract binding sites referred to as `glues' in place of binding domains with sequences: after sequence design, `matching' glues correspond to domains with complementary sequences.  Edges of the shapes used a special `null glue' with no valid binding partner.  In total, this initial design had 2,706 glues, and 1,456 tiles. 

The three shapes were then processed through a `merging' algorithm that attempted to reuse the same tiles in different shapes.  Each step of the algorithm randomly chose two tiles in two different shapes, with null glues on the same sides of each tile, if any.  It then considered a modified set where the two tiles were identical, by making them use the same four glues, and propagating the changes in the glues to all other places they occurred within all shapes, starting with the neighboring tiles (e.g., \efigref{extfig:merging}[c]).  Such a change could create undesired growth pathways, for example, allowing chimera of multiple shapes. Thus, the algorithm then checked the modified set for two criteria taken from algorithmic self-assembly (\efigref{extfig:merging}[ab]).  The self-healing criterion requires that, for any correct subassembly of any shape, while attachments of the wrong tile for a particular location may take place by one bond, only the correct tile can attach by two or more bonds\autocite{winfree_self-healing_2006}.  The second-order sensitivity criterion for proofreading requires that, for any correct subassembly of any shape, if an incorrect attachment by one bond takes place, the incorrectly attached tile will not create a neighborhood where an additional incorrect tile can attach by two bonds, and thus the initial error will be likely to fall off\autocite{winfree_proofreading_2004,evans_optimizing_2018}.  If the modified set satisfied these two criteria, which are trivially satisfied when every tile and bond is unique to a particular location, then the merging algorithm accepted the modified set, and continued to another step with a different pair of randomly-chosen tiles.  Thus, we ensured that there is at least a minimum barrier to continued incorrect growth in a regime where tile attachment by two or more bonds is favorable, and attachment by one bond is unfavorable, which is the case close to the melting temperature of most DNA tile assembly systems\autocite{winfree_simulations_1998,evans_physical_2017}.

The algorithm repeatedly merged tiles that satisfied the two criteria until no further acceptable merges were possible.  As each merge could affect the acceptability of later merges by changing the glues around each tile, in order to guide the algorithm toward a sequence of merges more likely to be compatible, the algorithm was initially restricted to considering pairs of tiles from an alternating `checkerboard' subset, which, apart from edges, were likely to be merge-able. After exhausting acceptable merges from this subset, the algorithm attempted merges using all tiles in the system.  After repeating this stochastic algorithm multiple times, and selecting the system with the smallest number of tiles, the final resulting system had 698 binding domain and 917 tiles, with 371 of tiles shared between at least two shapes (\efigref{extfig:merging}[d]).

After the assignment of abstract binding domains to each tile by the merging algorithm, the sequences for the binding domains, and thus tiles themselves, were generated using the sequence design software of Woods and Doty et al\autocite{woods_doty_diverse_2019}.  Tiles used a standard SST motif, with alternating 10 and 11 nt binding domains, designed to have similar binding strengths as predicted using a standard thermodynamic model\autocite{SantaLucia:2004dna,zadeh2011nupack,woods_doty_diverse_2019}. Following Woods and Doty et al\autocite{woods_doty_diverse_2019}, we set a target range of -8.9 to -9.2~kcal/mol for a single domain at 53\,\degC, which was between the melting temperature and growth temperature for their system.  Null binding domains on the edges of shapes, not intended to bind to any other tiles, were assigned poly-T sequences.

\subsection*{Models of nucleation.}
To model the dependence of the nucleation rates of the three shapes on patterns of unequal concentration, we developed a simple nucleation model based on the stochastic generation of possible nucleation pathways and critical nuclei.  The model estimates nucleation rates by analyzing stochastic paths generated in a greedy manner by making single-tile additions starting from a particular monomer in the system. At each step, all favorable attachments are added and then an unfavorable attachment is performed with probability weighted by the relative free-energy differences of the available tile attachment positions. When multiple favorable attachments are available, the most favorable attachment is made deterministically. This procedure is repeated for many paths over all possible initial positions within the shape considered, and the barrier (highest free energy state visited in `growing' a full structure) is recorded for each path. A nucleation rate is estimated by assuming an equilibrium occupation of this barrier state (Arrhenius' approximation \autocite{frenkel}) and summing over the kinetics of the available attachments from this state. See \efigref{extfig:nucleation} and \sisgm for a detailed discussion.  The approximations here could be improved by running fully reversible simulations, e.g., using {\tt xgrow} and the kinetic Tile Assembly Model\autocite{xgrow,winfree_simulations_1998} augmented with Forward Flux Sampling \autocite{Allen2005-ve}.

\subsection*{Fluorophore labels and DNA synthesis.}
Sites for fluorophore and quencher modifications were chosen to avoid edges, modify only unshared tiles, and provide a reasonable distribution of locations on each shape.  Fluorophores were chosen for spectral compatibility and temperature stability\autocite{you_measuring_2011}.  ROX, ATTO550, and ATTO647N were paired with Iowa Black RQ, and FAM was paired with Iowa Black FQ.  Both fluorophore and quencher modifications were made on the 5' ends of tiles; to sufficiently colocalize fluorophores and quenchers, one tile in the label pair used a reversed orientation (\figref{fig:flagscan}[a]).  Fluorophore labels are discussed in detail in \sifluorophores.

Tiles without fluorophore or quencher modifications were ordered unpurified (desalted) and normalized to 400~\muM in TE buffer (Integrated DNA Technologies).  Tiles with fluorophore or quencher modifications were ordered HPLC-purified and normalized to 100~\muM.  Given that unpurified synthetic oligonucleotides typically have less than 40\% to 60\% of the molecules being full length, it is remarkable (though consistent with Woods and Doty et al\autocite{woods_doty_diverse_2019}) that this did not prevent successful pattern recognition by nucleation.

\subsection*{Experimental overview.}
The basic workflow for the main experiments was as follows: For a chosen set of concentration patterns (flag or image), samples were prepared on a 96-well plate using an acoustic liquid handler to mix strand stocks in the necessary proportions; vortexed, spun, and transferred to PCR tubes for the days-long anneal in the qPCR machine; then samples were deposited on mica for AFM imaging.  Fluorescence from the qPCR machine and AFM images were subsequently analyzed.

\subsection*{Mixing and growth.}
Individual tiles were mixed, in the concentration patterns used for experiments, using an Echo 525 acoustic liquid handler (Beckman Coulter). Samples used TEMg buffer (TE buffer with 12.5~mM~MgCl$_2$) in a total volume of $\sim$ 20~\muL.  Flag experiments used a 50~nM base concentration of unenhanced tiles, and an 880~nM concentration of enhanced concentration tiles, while pattern recognition experiments employed tiles with nominal concentrations between 16.67~nM to 450~nM, which were then quantized into ten discrete values to simplify mixing and conserve material (see \sipatmix).  

For each concentration pattern in the flag experiments and pattern recognition of trained images, four samples were prepared, each with the same concentrations pattern of tiles, but with tiles in different locations replaced by their fluorophore/quencher-modified alternates: one sample for each shape with tiles for all four fluorophore labels on only that shape, to monitor growth of multiple regions on each shape, and an additional sample with one fluorophore on each shape: ROX, ATTO550 (`five'), and ATTO647N (`six') on H, A, and M structures respectively.  To reduce the total number of samples, only the lattermost sample type was prepared for pattern recognition of test images.  Fluorophore and quencher-modified tile locations always had tiles mixed at the lowest concentration used in the experiment.  \setbox0\vbox{\autocite{xgrow}\autocite{schulman_programmable_2009}}

After transferring samples to PCR tubes, samples were grown in an mx3005p quantitative PCR (qPCR) machine (Agilent), in order to provide a program of controlled temperature over time while monitoring fluorescence.  Growth protocols began with a ramp from 71\,\degC to 53\,\degC over 40 minutes to ensure any potentially preexisting complexes were melted, and then a slower ramp from 53\,\degC to an initial growth temperature at 1\,\degC per hour.  At this point, three different protocols were used.  For constant temperature flag growth experiments, the growth temperature was 47\,\degC, and this was held for 51 hours.  For temperature ramp flag growth, the initial growth temperature was 48\,\degC, which was reduced over 100 hours to 46\,\degC.  For pattern recognition, a ramp from 48\,\degC to 45\,\degC over 150 hours was used.  For constant temperature experiments, fluorescence readings were taken every 12 minutes, and for other experiments, every 30 minutes.  After the growth period, temperature was lowered to 39\,\degC at 1\,\degC per 26 minutes.  See \sitprots for temperature protocols plotted as a function of time. 
The experimental timescales and temperatures were chosen not to test the potential speed of selective nucleation, but rather to provide robustness to unknown nucleation temperatures and to convincingly show that nucleation of incorrect structures is limited over long timescales.  Thus, on-target nucleation often took place during a comparatively short time and temperature in the experiment, with the remaining time spent either above the expected nucleation temperature, or waiting to observe potential off-target nucleation.  We also did not try to optimize the system's speed: the WTA mechanism suggests that significantly faster timescales are possible, and smaller assemblies would reduce the time needed for growth after nucleation.
Because of the small sample size and long experiment duration, great care to avoid evaporation was necessary.  Once protocols were finished, samples were stored at room temperature until ready for AFM imaging.

\subsection*{Imaging.} 
AFM imaging was performed using a FastScan AFM (Bruker) in fluid tapping mode directly after annealing was completed. In contrast to previous studies\autocite{wei_complex_2012,ke_three-dimensional_2012,ong2017programmable} where uniquely-addressed SST shapes were gel purified prior to imaging, we did not do so here; thus we were able to observe assembly intermediates. To achieve better images, two techniques were combined: sample warming to prevent nonspecific clumping of structures, and washing with Na-supplemented buffer to prevent smaller material, such as unbound, single DNA tile strands, from adhering to the mica surface.  Each sample was diluted 50x into TEMg buffer with an added 100~mM~NaCl, then warmed to approximately 40\,\degC for 15 minutes.  50~\muL of the sample mix was deposited on freshly-cleaved mica, then left for two minutes.  As much liquid as possible was pipetted off of the mica and discarded, then immediately replaced with Na-supplemented buffer again, and mixed by pipetting up and down.  This washing process of buffer removal and addition was repeated twice with added-Na buffer, then once with TEMg buffer to remove remaining Na, before imaging was performed in TEMg buffer.  As adhesion of DNA to mica is dependent upon the ratio of monovalent and divalent cations in the imaging buffer, this process was meant to ensure that unbound tiles were removed during the washing process where Na and Mg were present, while imaging itself took place with only Mg, so that the lattice structures would be more strongly adhered to the surface, resulting in better image quality.

\subsection*{Fluorescence and AFM data analysis.}
Fluorophore signals are known to be affected by extraneous factors such as temperature, pH, secondary structure, and the local base sequence near the fluorophore\autocite{you_measuring_2011}, which complicates quantitative interpretation of absolute fluorescence levels.
Our own control experiments also illustrated effects due to partial assembly intermediates as well as due to the total amount of single stranded DNA in solution (\sifluorophores).
For this reason, the fluorescence of each fluorophore was normalized to the maximum raw fluorescence value of that fluorophore in that particular sample, and the \emph{time} at which the fluorescence signal decreased by 10\% was then 
used as a measure of the extent of nucleation that appears less sensitive to these artifacts (\efigref{extfig:fluorophores}).
The duration between the point of 10\% quenching and the end of the growth segment of the experiment was defined as the `growth time' for that fluorophore label; the growth time was defined as 0 in the event of quenching never reaching 10\%.  For concentration patterns with four samples with different fluorophore arrangements, the total growth time of a shape was defined as the average of the growth time of the five total fluorophore labels on the shape across the four samples (four in the shape-specific sample, and one in the each-shape sample), while for concentration patterns with only one sample, the growth time of the corresponding fluorophore label was used.  As the position of the fluorophore within the shape, relative to where nucleation occurs, has a substantial influence on growth time measurements, the considerable variability in these measurements relative to the true nucleation kinetics must be acknowledged.

For flag experiments, AFM imaging was done only for qualitative confirmation of the selective nucleation and growth indicated by fluorescence results.  For pattern recognition and equal-concentration experiments, however, shapes in AFM images were uniformly quantified.  At least one sample of each of the patterns had three 5 $\times$ 5 $\mum$  images taken under comparable conditions.  The sample corresponding with each image was blinded, and structures were counted independently by each of the four authors, classifying structures as either ``nearly complete'' or ``clearly identifiable'' examples of each of the three shapes.  For the purposes of analysing pattern-dependent nucleation and growth, no clear distinction between the number of nearly complete and clearly identifiable shapes was found, and so the two categories were summed.  Counts were averaged across the three images, then averaged across the counts of the four authors, to obtain a count per shape per 25 $\mum^2$ region for each pattern. 
Each author used their own, subjective, interpretation of ``nearly complete'' and ``clearly identifiable'' structures, and the total number of structures counted in each image differed by up to $\pm 50\%$ for different authors.  However, the ratios of different shapes in each image counted by each author remained within 5\% of the mean ratios for most images, and across all images, no author had a bias of more than $\pm 4\%$ toward identifying a particular shape more or less often than average.  Results are detailed in \sicountdata.

To measure the selectivity of patterns, the fraction of on-target shape growth time, and AFM counts, compared to the sum of shape growth times and AFM counts, was used.  The total growth times, and total AFM counts, of the on-target shapes were used to measure overall shape growth.

\subsection*{Pattern recognition training.}
Images for pattern recognition were selected from several sources, rescaled to 30 $\times$ 30, discretized to 10 grayscale values, and adjusted so that the number of pixels with each value was consistent across all images (see \siimages for details).  Each pixel's grayscale value, $0 \le p_n \le 1$, was converted to the concentration $c_i$ for the corresponding tile $t_i$ using an exponential formula, $c_i = c e^{3\;p_n\ln 3}$, where the base concentration is $c=16.67$~nM. The intention of the numbers used was to make the average tile concentration 60~nM for each image.  As each image had 900 pixels and there are 917 tiles in the system, 17 tiles did not have their concentrations set by any pixel; these tile concentrations were uniformly set to the lowest concentration, and the assignment of these tiles was used to ensure that fluorophore label locations did not vary in concentration.

The tile-pixel assignment was optimized through a simple hill-climbing algorithm, starting from a random assignment, where random modifications to the assignment map are attempted at each step and accepted if the move increases the efficacy of the map. This efficacy was quantified through a heuristic function that accounts for relative nucleation rates, location of nucleation sites (with preference given to locations that succeeded in the flag experiments shown in \figref{fig:flagscan}[d]), and satisfaction of constraints related to the fluorescent reporters. Because the nucleation algorithm described above is costly, a simplistic model of nucleation based upon the Boltzmann-weighted sum of concentrations over a $k \times k$ window swept over each structure (similar to the model employed in Zhong et al\autocite{zhong_associative_2017}) was used to evaluate relative nucleation rates for a majority of the optimization steps. The more detailed but computationally costly model described above was then employed for an additional several hours in hopes of improving the mapping.  The window-based nucleation model (along with all constraints about nucleation location and fluorescent reporters) is also employed to explore the capacity of this map training procedure in \efigref{extfig:capacity}.  Details of the pattern recognition training and the window-based nucleation model are discussed in \sialltraining. \phantom{\autocite{weibrecht2010proximity,schaus2017dna}}

\subsection*{Data availability.}
AFM images, fluorescence trajectories, DNA sequences, and simulation results are available at \url{https://www.dna.caltech.edu/SupplementaryMaterial/MultifariousSST/}.

\subsection*{Code availability.}
Algorithms for tile set design, sequence design, nucleation rate prediction, pixel-to-tile map optimization are available at
\url{https://www.dna.caltech.edu/SupplementaryMaterial/MultifariousSST/}.

\printbibliography[segment=2,check=onlynew]

\end{methods}

\begin{addendum}
	\item[Acknowledgements.]
	We thank Michael Brenner, Jehoshua Bruck, Aaron Dinner, David Doty, Deborah K. Fygenson,  Stanislas Leibler, Richard M. Murray, Lulu Qian, Paul W. K. Rothemund, Petr \v{S}ulc, Chris Thachuk, Grigory Tikhomirov, Damien Woods, and Zorana Zeravcic.  Ting Zhu, Thomas Ouldridge, Salvador Buse, Matthew Alexander, Mohini Misra, Anna Lapteva also provided valuable feedback on early drafts. We thank Zorana Zeravcic for assistance with artwork in Figure 1.
   \item[Funding.]
	   Supported by National Science Foundation grants CCF-1317694 and CCF/FET-2008589, 
	   the Evans Foundation for Molecular Medicine, 
	   European Research Council grant 772766, Science Foundation Ireland grant 18/ERCS/5746,
	   and the Carver Mead New Adventures Fund.
JOB, AM were primarily supported by the University of Chicago
Materials Research Science and Engineering Center, which is funded by National
Science Foundation under award number DMR-2011854. AM acknowledges support from the Simons Foundation.
   \item[Author Contributions.]
	   CGE, EW, AM conceived the study.
	   CGE and EW designed the molecules.
	   CGE, JOB, EW, AM wrote simulation code, designed the experiments and performed the experiments, analyzed the data and wrote the manuscript.
   \item[Competing Interests.] 
	   The authors declare that they have no competing financial interests.
   \item[Correspondence.] 
	   Correspondence should be addressed to \texttt{cge@dna.caltech.edu},
	   \texttt{jdobrien07@gmail.com}, 
	   \texttt{winfree@caltech.edu}, or \texttt{amurugan@uchicago.edu}.
   \end{addendum}
   
\renewcommand\thefigure{E\arabic{figure}}    
\setcounter{figure}{0}

\begin{figure*}[ht]
\centering
\includegraphics[width=6.0in]{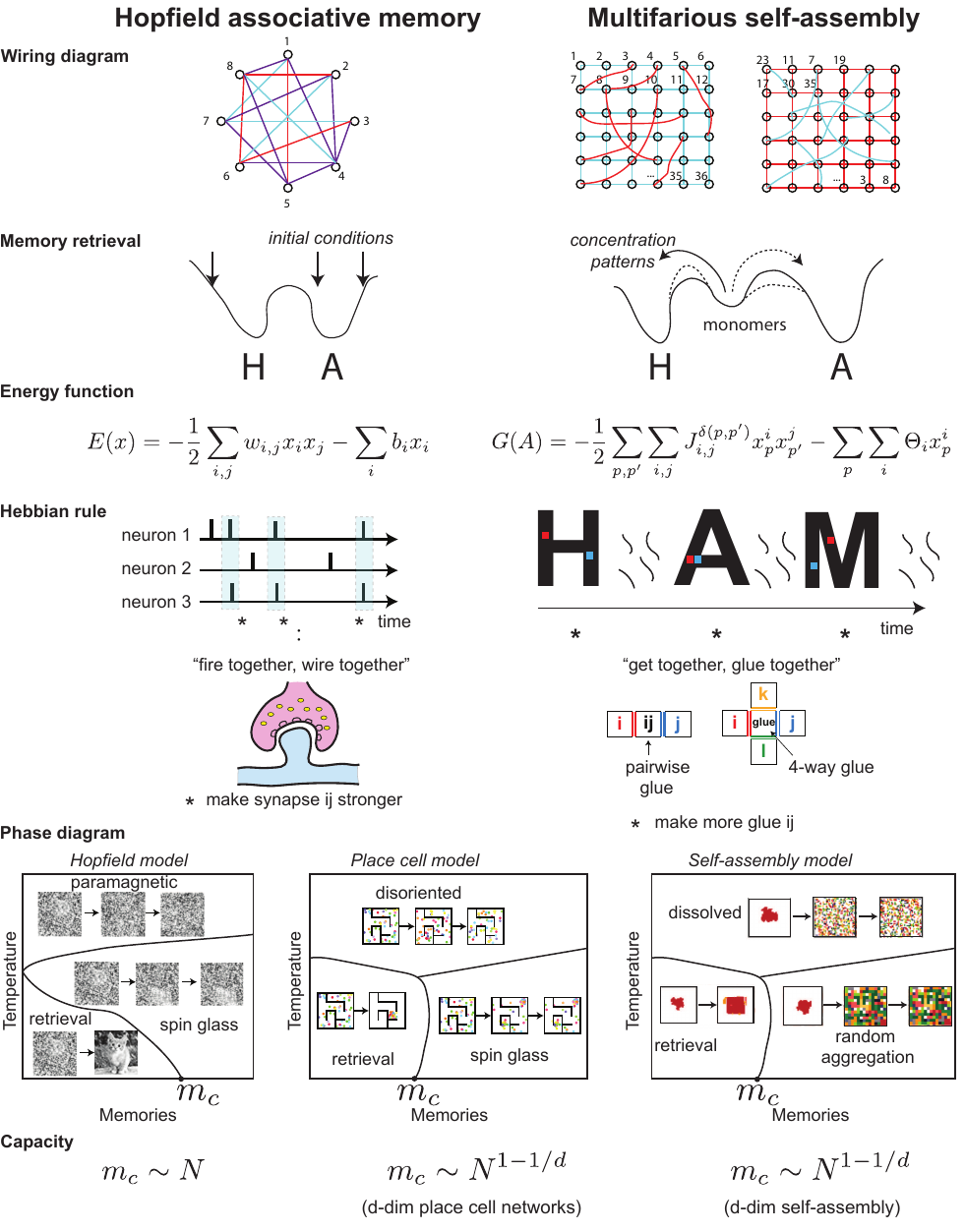} \vspace{-.3cm}
\caption{\textbf{Parallels and differences between neural network models and self-assembly models as exemplars of collective behavior.}
In this rough metaphor, a neuron corresponds to a tile.  While Hopfield networks allow full connectivity, multifarious self-assembly (like place cell networks) restricts connectivity to a superposition of grids with different unit permutations.  The state of a Hopfield network consists of the set of active neurons, while the state of an assembly consists of the set of tiles present and their arrangement, which is restricted to be connected.
We use $x_i \in \{-1,+1\}$ for the activity of neuron $i$, and $x^i_p \in \{0,1\}$ for the occupancy of tile $i$ in position $p$.
The energy of a state is a quadratic function governed by synaptic weights $w_{i,j}$ and biases $b_i$ for neural activities, or for assemblies, by directional binding energies $J^\delta_{i,j}$ for tiles $i$ and $j$ at positions $p$ and $p'$ that are neighbors in direction $\delta$, along with (inverted) tile chemical potentials $\Theta_i$.
An environment presents a sequence of outside influences driving system state, either stimulating neural activity or spatially organizing tiles.
Learning in Hopfield networks occurs any time neurons are simultaneously active. For self-assembly, learning an interaction requires tiles $i$,$j$ to be located next to each other; we envision a hypothetical proximity-based ligation process\autocite{weibrecht2010proximity,schaus2017dna} that creates interaction mediating glues $ij$ for molecules $i$,$j$ that spend time together in spatial proximity.
Qualitative system behaviors depend on the number of memories being stored and the operating temperature, including phases where system state randomizes (paramagnetic / disoriented / dissolved), gets locked in a spurious local minimum (spin glass / random aggregation), or successfully retrieves learned memories.
Due in large part to the restrictions on connectivity, the capacity of place cell networks and multifarious self-assembly is less than for the Hopfield model.
See \sineural for details and discussion.}
\label{extfig:neural}
\end{figure*}

\begin{figure*}[ht]
\centering
\includegraphics{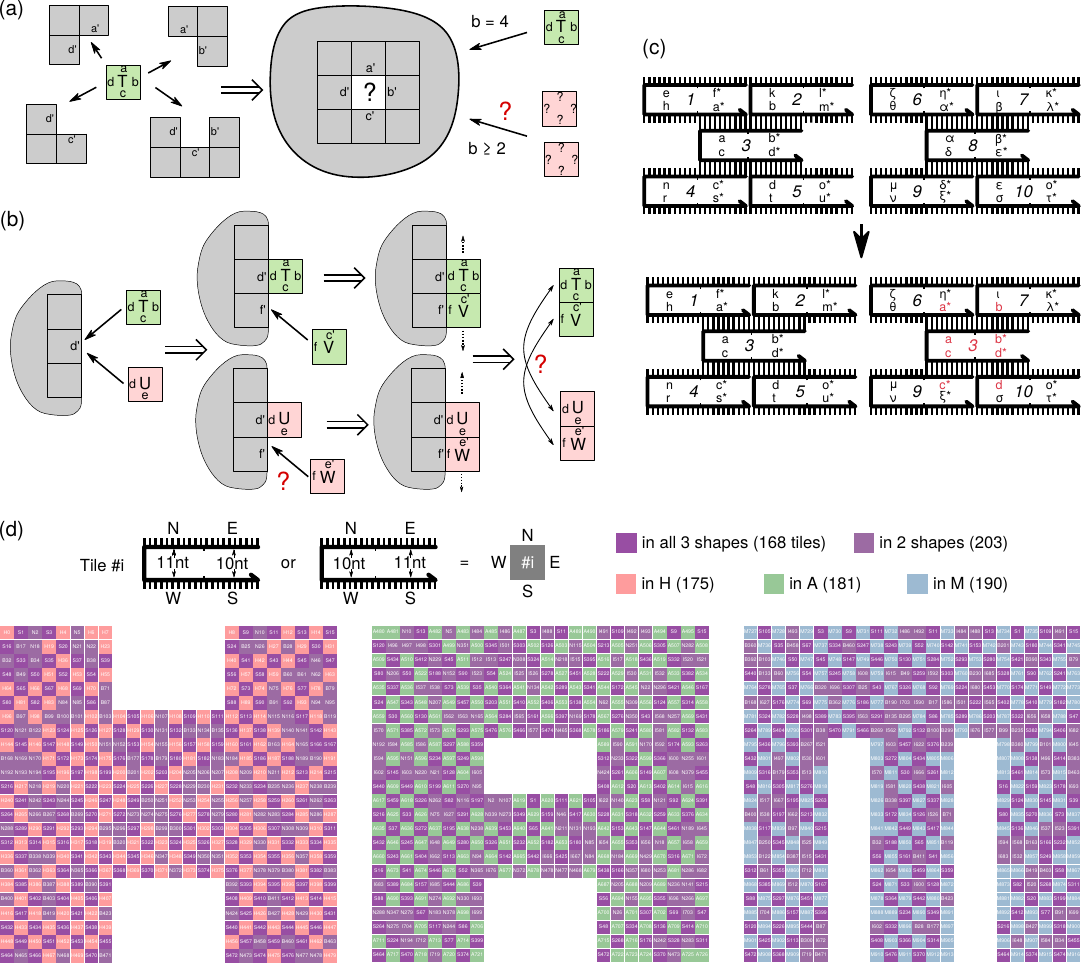}
\caption{\textbf{Proofreading tile set design and tile assignment map.}
Extensive promiscuous interactions present in the SHAM mix could in principle lead to unintended chimeric structures and other malformed assemblies.  To reduce or prevent such behaviors, 
our design incorporates self-assembly proofreading principles, so called because they enhance quick rejection of mis-assembled tiles. Much like with neural networks\autocite{Hopfield2010-lf}, random arrangement of tiles (such as the initial checkerboard layout in the first stage of our design process) provides a statistical proofreading\autocite{Murugan2015-ps} in the sense that problematic interactions are unlikely to arise.  
Further optimization of the tile set (in our second stage) ensures that two types of problematic interactions do not occur, thereby conferring algorithmic proofreading\autocite{winfree_proofreading_2004}
and self-healing properties\autocite{winfree_self-healing_2006}.
This tile set optimization is derived from prior work\autocite{evans_optimizing_2018}. 
\textbf{a,} Our systems are designed to grow in a regime where a tile attaching by at least two bonds is favorable, but a tile attaching by one bond is not (`threshold 2').   Motivated by self-healing tile systems~\autocite{winfree_self-healing_2006}, we seek a tile set where no correct partial assembly should ever allow an undesired tile to attach by two or more bonds, though undesired attachments by one bond are allowed, such that any favorable attachment to a partial assembly will be correct. 
\textbf{b,} In addition to tiles attaching favorably by $2$ bonds to growing facets, new facets in the system will only be created by tiles attaching unfavorably by one bond, and then being stabilized by further, favorable growth.  
At a site where tile $T$ would correctly attach by one bond, a tile $U$ might be able to attach incorrectly by the same bond.  $T$ would correctly be stabilized by the subsequent attachment of $V$ by two bonds, but $U$ might be stabilized as well if there is a tile $W$ that can attach to it and shares the same glue as $V$.  
Thus, if for every pair of tiles that can bind to each other (e.g., $T+V$), there is no other pair of binding tiles (e.g., $U+W$) that share two glues on the same edges of the tiles, then any tile that attaches by one bond to an assembly will either be the correct tile, or will not allow a subsequent stable attachment, and will likely detach quickly.
This is equivalent to `second-order sensitivity' with all directions treated as inputs, functioning as a form of self-assembly proofreading~\autocite{evans_optimizing_2018,winfree_proofreading_2004}.
\textbf{c,} We created a multifarious tile system by first starting with three shapes constructed entirely of unique tiles, then repeatedly attempting to `merge' tiles in different shapes by constraining the sequences of their domains to be identical, and checking whether each merge of two tiles results in a tile system that does not have any tile pairs violating criteria in \textbf{a} and \textbf{b}. \textbf{d,}
From multiple trials of the merging process, each initially favoring a checkerboard arrangement before attempting more general merges, we selected the smallest result containing 917 tiles.
DNA sequences for tiles in the system were designed with the single-stranded tile (SST) motif\autocite{yin_programming_2008}, with two alternating tiles motifs of 10~nt and 11~nt domains (full shape layouts and tile sequences are shown in \sitiles).}
\label{extfig:merging}
\end{figure*}

\begin{figure*}[ht]
\centering
\includegraphics{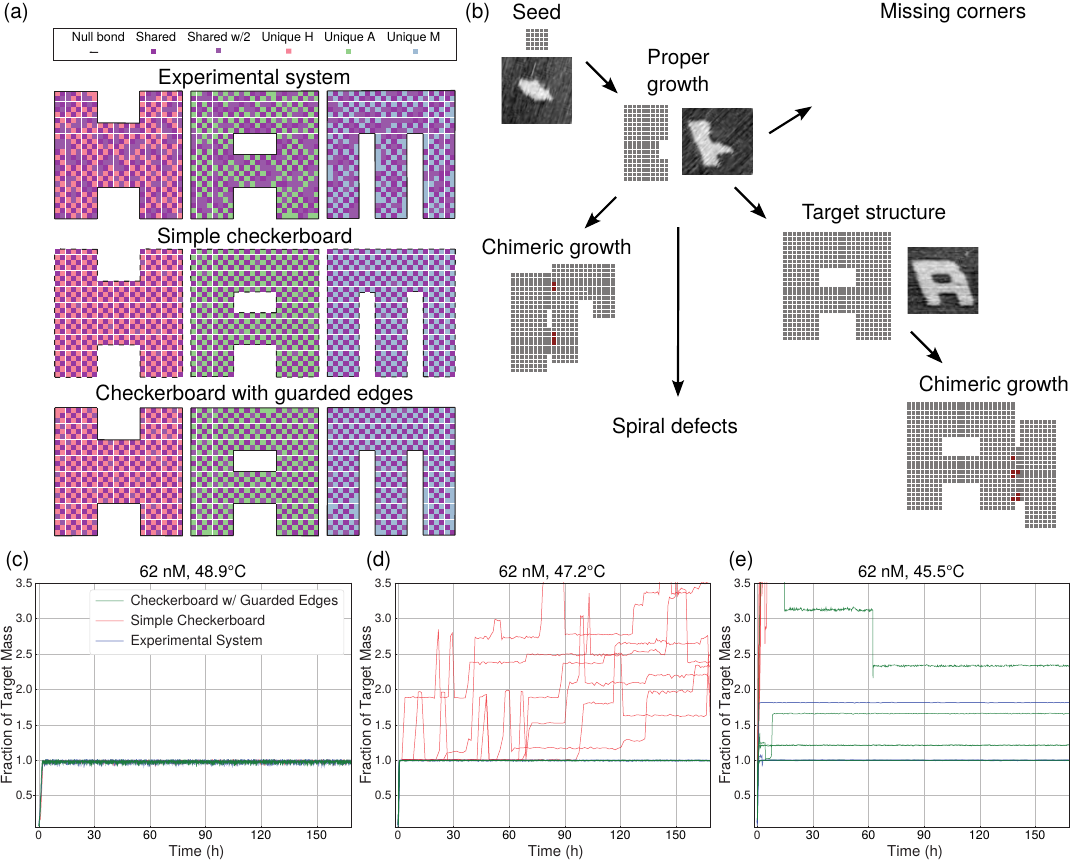}
\caption{\textbf{Suppression of chimeric growth through tile set design.}
\textbf{a,} We use simulations to contrast assembly errors in three distinct tile sets: the proofreading tile set with an inert boundary used in experiments, described in \figref{fig:multifarious} \textbf{(a, top)}; 
a simple checkerboard tile set with a strictly alternating shared and unique tile pattern for each shape, where 
unique tiles can be seen as mediating different interactions between shared tiles \textbf{(a, middle)}; and an edge-guarded checkerboard in which we additionally enforce inert bonds around each shape's perimeter \textbf{(a, bottom)}.
For each tile set, we performed kinetic growth simulations, starting from a pre-formed $5\times 5$ seed taken from a location within H.
Simulations were performed using the kinetic Tile Assembly Model as implemented by {\tt xgrow} (with chunk fission) \autocite{xgrow} with uniform tile concentrations corresponding to 62 nM and parameters estimated in \siktam.
\textbf{b,} Schematic illustrates various desired and undesired growth pathways for A, along with representative AFM images taken from the A flag 1 experiment
(\siAflag1).  Two distinct kinds of chimeric structures were seen in simulation as the result of promiscuous interactions: chimeric structures can grow either before full assembly of the target structure (e.g., part-A, part-M) or emerge spontaneously from the edge of a properly formed structure (e.g. full-A, part-H). Chimeras like those illustrated along the lower path are held together by just a few bonds and sometimes can quickly break apart (tiles with unintended bonds are shown in red); these result in sharp drops in simulated assembly size, as the simulation discards one subassembly when disconnected. 
Note that chimeric growth was not observed experimentally, possibly as a result of effective experimental system design; however, many observed structures failed to complete the upper right and/or lower left corners, or appeared to have suffered a spiral growth defect.
A possible explanation for the missing corners, which is also seen in H and M, is supported by coarse-grained molecular dynamics simulations of SST lattice curvature (\sioxDNA).  Spiral defects were not seen in H or M and are presumably due to the interior hole in A.
\textbf{c--e,} The size of the assembly (in units of the size of the fully formed H) is shown as a function of time. 
For higher temperature 48.9\,\degC (\textbf{c}), no chimeras are observed on the simulated timescales for any tile set. For intermediate temperature 47.2~\degC (\textbf{d}), all 6 checkerboard trajectories still result in chimeras, while no errors are observed on the timescale probed for the guarded checkerboard or experimentally-implemented proofreading tile set. 
For lower temperature 45.5\,\degC (\textbf{e}), chimeras are seen in all runs for checkerboard structures (red traces), 4 of the 6 runs for guarded checkerboard structures (green traces) and 1 of the 6 runs for proofreading structures.
}
  \label{extfig:errorsims}
\end{figure*}

\begin{figure*}
    \centering
    \includegraphics{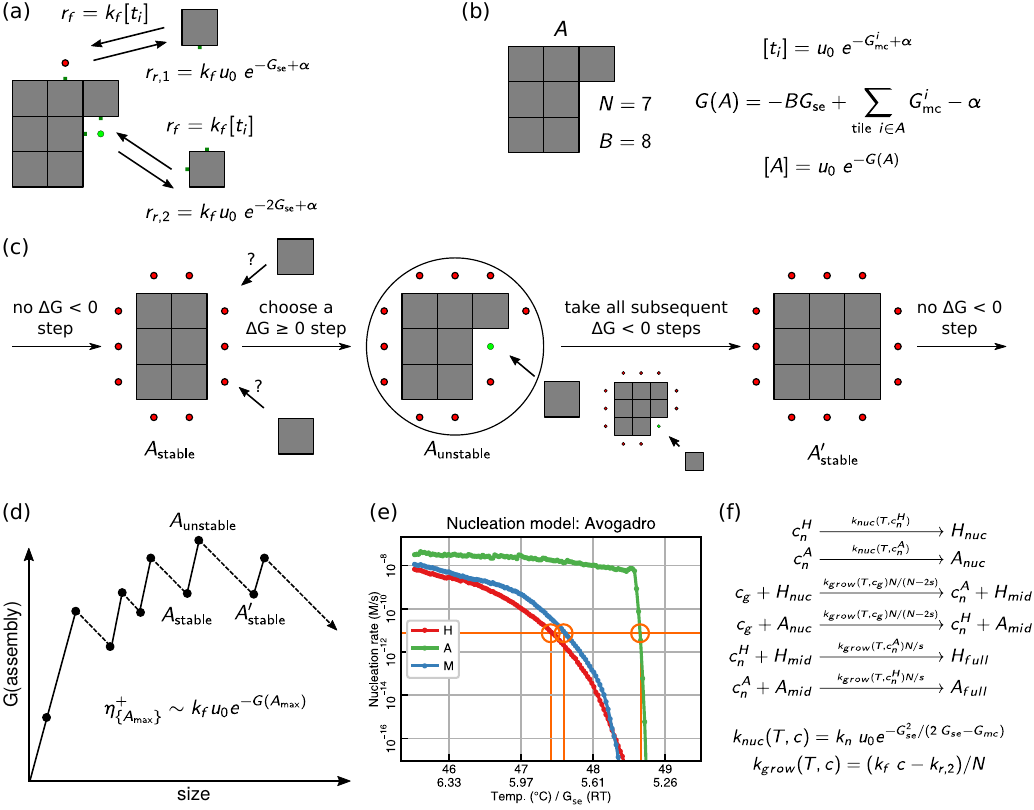}
    \caption{%
\textbf{Stochastic Greedy Model of nucleation, based on repeated stochastic simulations.}
\textbf{a,} The frequently-used kinetic Tile Assembly Model (kTAM)~\autocite{winfree_simulations_1998,evans_physical_2017} has rates for tile attachment and detachment events based on tile and assembly diffusion and total binding strength of correct attachments a tile can make at a lattice site.  Here $u_0 = 1$~M.
\textbf{b,} These rates can be used to derive a free energy for any tile assembly in a system, and, assuming fixed monomer concentrations, an equilibrium concentration for any assembly. Schulman \& Winfree \autocite{schulman_programmable_2009} showed that the equilibrium concentration of the highest-energy assembly along a nucleation trajectory under this assumption provides an upper bound for nucleation rate through that trajectory, with or without fixed monomer concentrations.
However, in a large system, considering all possible intermediate assemblies and all pathways, including many that are extremely unlikely, would be infeasible.  
Thus, we developed the Stochastic Greedy Model (SGM) to generate stochastically-chosen paths of tile attachments.  
\textbf{c,} Starting from a single tile (chosen with probability proportional to relative concentration), whenever the assembly is in a state $A_\mathrm{stable}$ where there is no tile attachment that would be favorable (have $\Delta G < 0$), one of the possible unfavorable (with $\Delta G \geq 0$) attachments is stochastically chosen, resulting in a higher-$G$ state $A_\mathrm{unstable}$. Then, all subsequent possible $\Delta G < 0$ attachments are made, resulting in the next $A'_\mathrm{stable}$ state; for our system of unique tiles for each site in the lattice, this sequence of favorable steps has a unique resulting assembly.  
\textbf{d,} The process repeats until all tiles in a shape are attached, which results in a trajectory with a maximum-$G$ assembly that can be used to bound the rate of nucleation, $\eta$, through that particular trajectory.  
\textbf{e,} By using this process to collect many trajectories, and then repeating the entire process for each of the three shapes in the system, we can estimate nucleation rates dependent upon temperature, with the assumption that tile monomer concentrations do not deplete, and that the trajectories found are a reasonable representation of likely trajectories.
For comparison between model predictions and experimental data in \efigref{extfig:flags}[d] and \ref{extfig:patterns}b, 
we determined the temperature at which the model predicted the nucleation rate exceeded a threshold (orange line), to compare with when fluorescence quenching exceeded a threshold. 
For details on the SGM model, see \sisgm.
\textbf{f,} To study the winner-take-all effect, we use a simplified chemical reaction network (CRN) model for the case of systems with shared tiles (shown here) and a similar model for systems without shared tiles (described in \siwtacrn).  Here, $c_n^H$ represent tiles in the flag area of shape H, which have initially higher concentrations; $c_n^A$ are the corresponding tiles in the flag area of shape A, which have normal concentrations; and $c_g$ represent tiles involved in growth from the nucleated seed $H_{nuc}$ to the almost-complete structure $H_{mid}$; and similarly for structure $A$.
A more detailed model based on (but simpler than) the SGM gives qualitatively similar results, as detailed in \siwtacrn.
}
\label{extfig:nucleation}
\end{figure*}

\begin{figure*}
    \centering
    \includegraphics{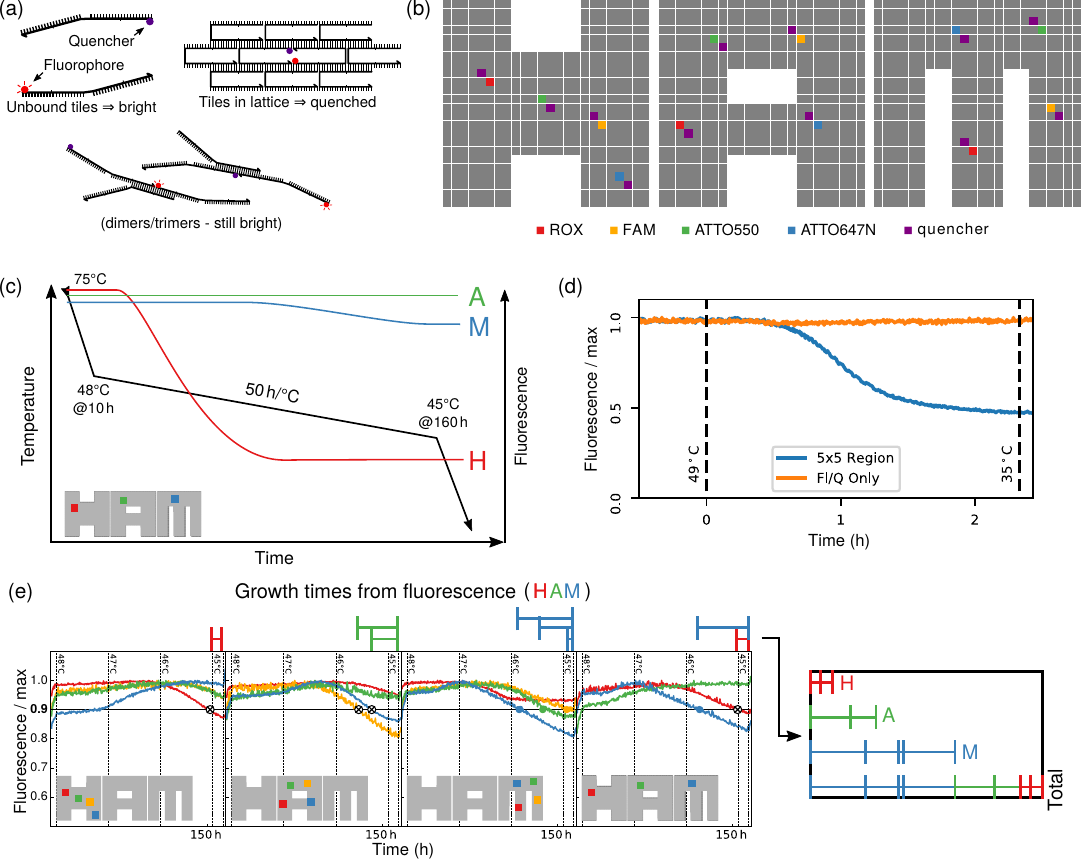}
    \caption{\textbf{Fluorophore quenching as a measure of nucleation and growth.}  
    \textbf{a,} Fluorescent labels used a fluorophore-quencher pair placed on the $5'$ ends of two modified tiles unique to one shape, where they were colocated, but had no complementary binding domains, ensuring that dimers could not form, and trimers would not closely colocate the fluorophore and quencher.  To constrain the pair to be close enough to quench in a well-formed lattice, one of the two tiles had its orientation and crossover position swapped compared to the unmodified tile for the location.
    \textbf{b,} Positions and types of all fluorophore/quencher pairs available for use.  For one sample, one position for each of four types of fluorophores could be chosen, and tile pairs for those locations replaced by their modified counterparts.  Thus different samples could probe different arrangements of up to four locations; four arrangements were used in experiments (e.g., in \textbf{e}). 
    \textbf{c,} Expected behavior of fluorophore labels on shapes as one shape nucleates and grows.  
    \textbf{d,} Fluorescence data for non-quenching (fluorophore tile only, orange) and quenching ($5\times 5$ lattice around fluorophore and quencher tiles, blue) controls for the ATTO647N fluorophore/quencher pair on A. Here, the temperature ramps linearly from $49\,\degC$ to $35\,\degC$  at a rate of $0.1\,\degC/\text{min}$, with all tiles at 50~nM, and each sample has its fluorescence normalized to its maximum value independently.
    \textbf{e,} An example of fluorescence growth time measurements (Mockingbird; see \simockingbird).
    Each fluorophore signal, in each sample, is independently normalized to its maximum value during the experiment, and the time between the point where the signal goes below 0.9 (``10\% quenching'') and the end of the experiment is measured (``growth time'').
    These times are then summed for all fluorophores, in all four samples, on each shape, resulting in a growth time for each shape, and, when normalized to the sum of all growth times, a relative growth time for each shape.
    See Methods and \sifluorophores\ for design and characterization of the fluorescence readout method, as well as an estimate of the melting temperature of H.
    }
    \label{extfig:fluorophores}
\end{figure*}

\begin{figure*}
    \centering
    \includegraphics{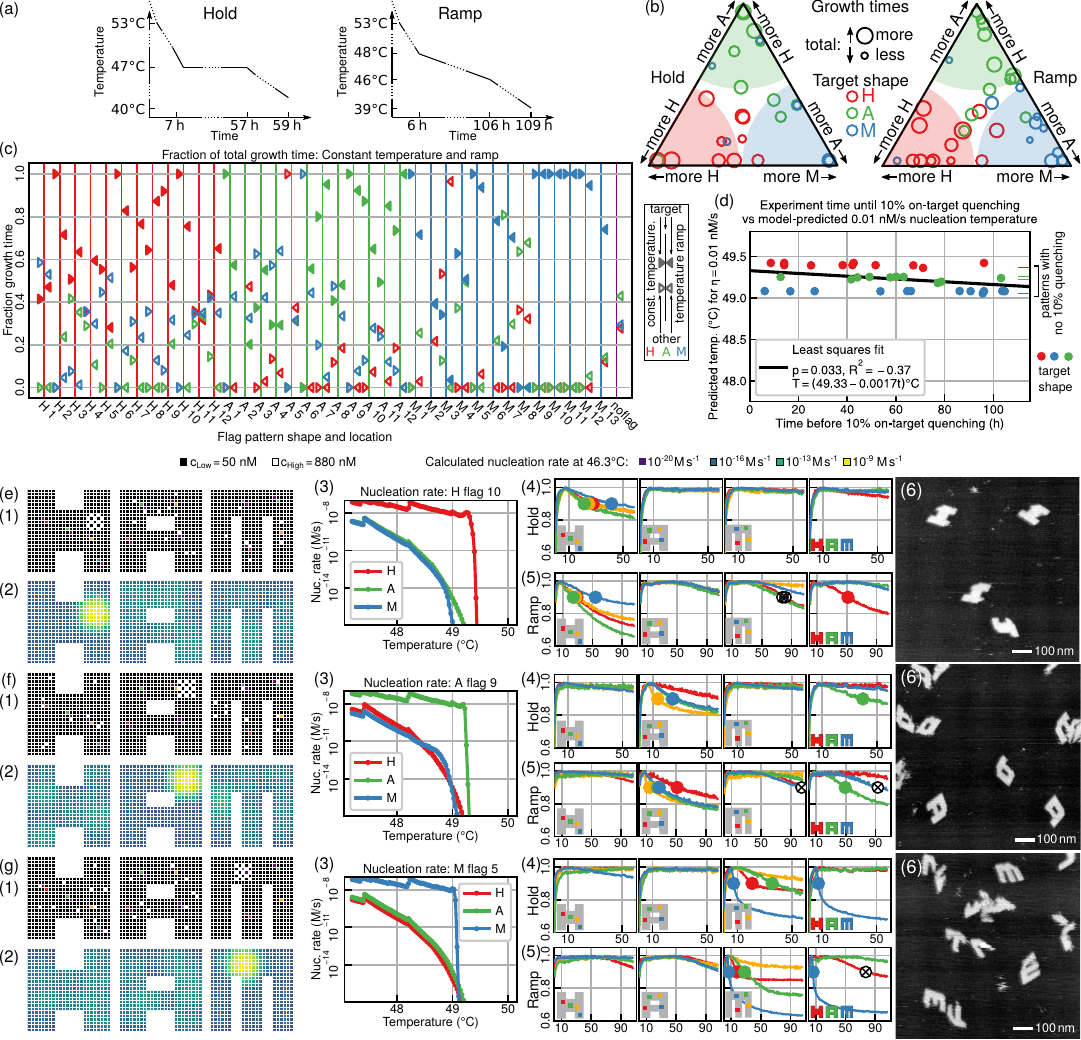}
    \caption{\textbf{Nucleation and growth with `flag' patterns of enhanced concentration.}
    \textbf{a--c,} 37 different concentration patterns with enhanced concentrations of shared tiles in $5\times 5$ regions were prepared, each with four different standard sets of fluorophores in four samples, and grown using two temperature protocols (\textbf{a}): a ramp focusing on $48\,\degC$ to $46\,\degC$ over 100 hours, and a hold at $47\,\degC$. 
    Using growth times as described in \efigref{extfig:fluorophores}, fluorescence data for many samples in both experiments showed preference for the desired shapes (\textbf{b, c}), but with considerable variation in selectivity and total amount of growth.  
    \textbf{d,} No statistically significant correlation was found between the nucleation model prediction for temperature of on-target nucleation and the time of on-target shape quenching in the temperature ramp experiment. 
    Although the nucleation model overestimates the nucleation temperature and its nucleation rate estimates may be far off, our interest here is in the qualitative features and difference between the shapes.
    \textbf{e--g,} Details of three patterns, with concentration patterns (1), weighted critical nucleus free energy starting from particular tiles (2), nucleation-model-estimated nucleation rates (3), temperature hold (4) and temperature ramp (5) experiment fluorescence results, and (6) AFM images from the temperature hold experiments. Information for all individual flag patterns is available in the \siflagdata. 
    }
    \label{extfig:flags}
\end{figure*}

\begin{figure*}[ht]
\begin{center}
\includegraphics{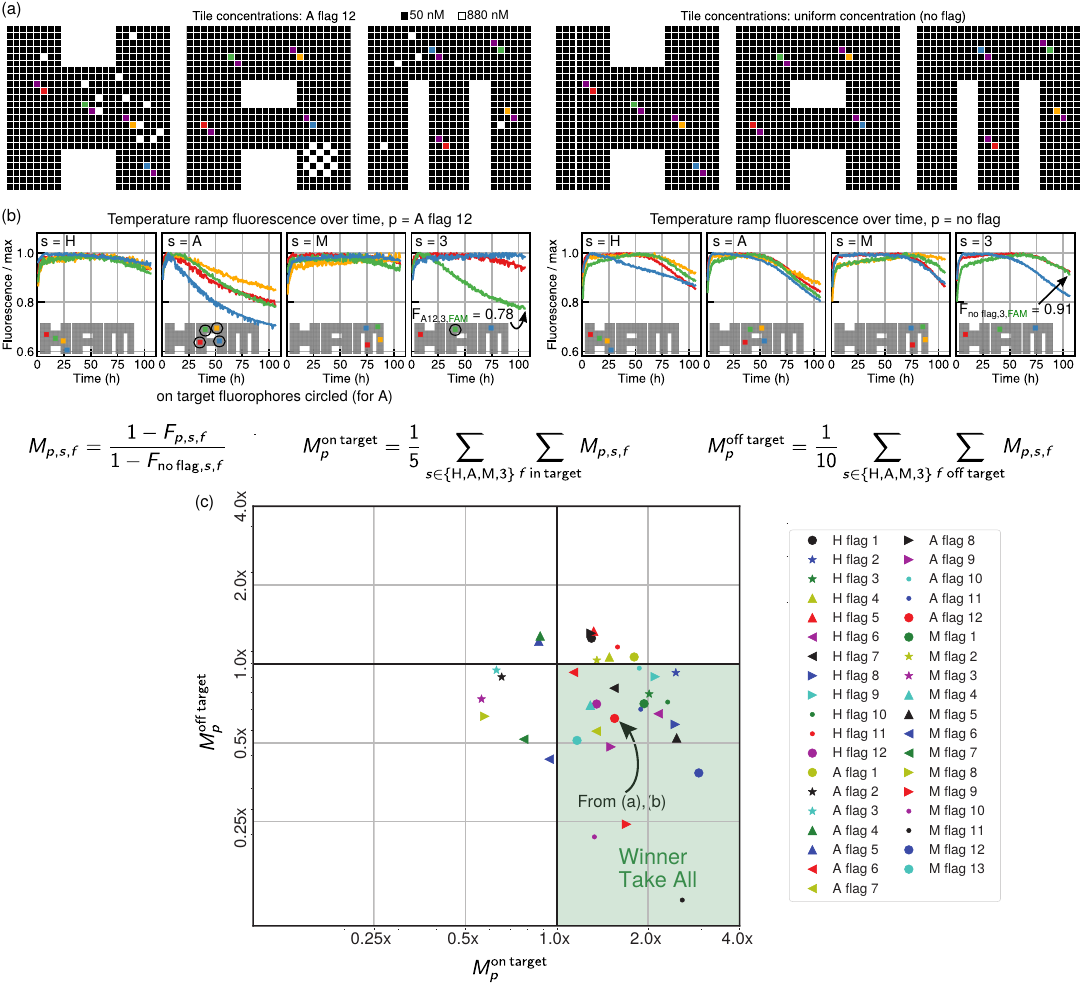}
\caption{\textbf{Evidence of winner-take-all in flag experiments.} \textbf{a,} An example flag pattern (A flag 12), and uniform 50 nM concentration `no flag' pattern. \textbf{b,} Fluorescence normalized to maximum readings, from the temperature ramp experiment (\efigref{extfig:flags} and \siflagdata).  The fluorescence at the end of the experiment, $F_{p,s,f}$, of fluorophore $f$ in sample $s$ of pattern $p$ is used along with the corresponding fluorescence value for the no flag pattern, $F_{\mathsf{no\, flag},s,f}$, to calculate the ratio $M_{p,s,f}$.  This ratio corresponds to the relative amount of quenching for that fluorophore in the flag pattern compared to the no flag pattern.  The ratios are averaged across the 5 on-target fluorophores (circled in \textbf{b}) in samples for the flag pattern to obtain an average on-target ratio, and across the 10 off-target fluorophores to obtain an average off-target ratio.  \textbf{c,} The on- and off-target ratios are plotted for each flag pattern.  For winner-take-all behavior, on-target quenching is expected to be higher with a flag pattern than with no flag, resulting in $M_p^{\mathsf{on\,target}} > 1$, while off-target quenching is expected to be reduced, resulting in $M_p^{\mathsf{off\,target}} < 1$. \label{extfig:wta}}
\end{center}
\end{figure*}

\begin{figure*}
    \centering
    \includegraphics{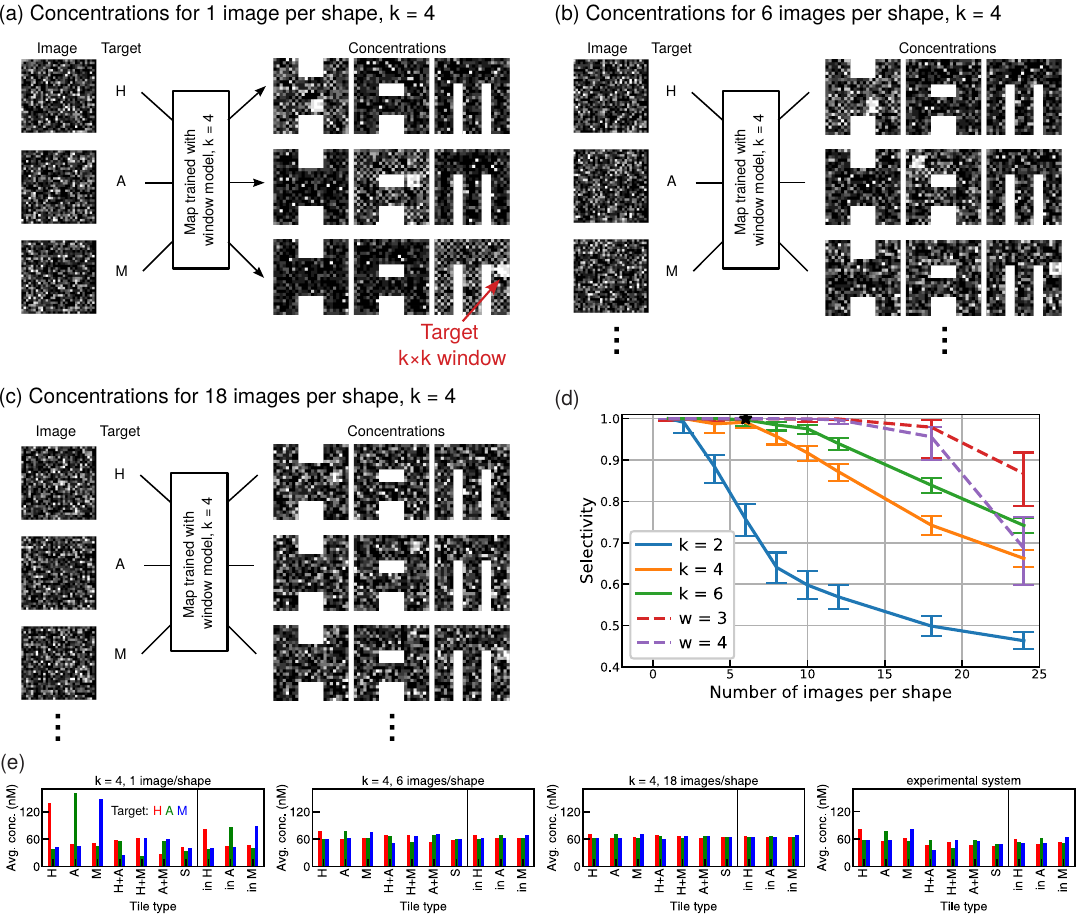}
    \caption{\textbf{Pattern recognition capacity.} To analyze the pattern-recognition capabilities of the designed tile set, the map-training algorithm (see \simaptraining) was run for increasingly larger sets of random images. 
    \textbf{a-c,} Example images mapped to concentration patterns for sets with 1, 12, and 18 trained images per shape, with the intended target shape for each image indicated.  Following the same procedure as used for the experimental system, with the same weighting of locations, $30\times 30$ images with 10 possible grayscale values and matching histograms were mapped exponentially to tile concentrations in the 917 tile system; however, all images were generated randomly.  Training was done using only the Window Nucleation Model with a window size $k$ of either 2, 4, or 6, with a limit of 400,000 steps (\siwindow).  For each number of images per shape considered, ten repetitions of training (starting from random assignments) were performed (to account for variability of the training algorithm) for each of three different sets of images (to account for variability in sets of images).
     \textbf{d,} As the number of images in the set increases, the selectivity of nucleation using the trained map decreases. For larger $k$, the pixel-tile map can exploit higher-order correlations and can thus accommodate more images.  For each fully-trained system, nucleation rates were calculated using the Stochastic Greedy Model, described in \sisgm, at $\gse=5.4$, which roughly corresponds to a temperature of 48.6\,\degC, and with concentrations comparable to the experimental system.  Selectivity was calculated as the nucleation rate of the target shape for each image divided by the total nucleation rate of all three shapes for that image, averaged over all images in the system, and over all 30 systems (10 repetitions for each of 3 sets of images) for each point, with 90\% confidence intervals shown.  Star shows selectivity calculated from nucleation model results for the experimentally-implemented system.
    Alternatively, dashed lines show results (at $\gse=5.5$) for maps constructed by a simpler training method that assigns the highest \(w^2\) previously-unassigned pixels in each training image to a unique $w\times w$ region in the target shape, detailed in \sisimplertraining.  These maps have at least as much capacity as the model-trained maps within the time constraints of these tests, suggesting a robustness to training method.
    \textbf{e,} As the number of images increases, pattern recognition must increasingly rely on patterns of concentrations of shared tiles, rather than choosing a pixel-to-tile map that places high-concentration pixels on tiles unique to the target shape.  Histograms show average concentrations of tiles in different shapes or combinations of shape
    (including the average across tile categories)
     for images in training cases \textbf{a--c}, and the experimental system.  The change can also be seen in the concentration maps of \textbf{a--c}, with the sharp checkerboard of high concentration tiles in target shapes in \textbf{a} becoming less apparent in \textbf{b} and \textbf{c}.
     }
    \label{extfig:capacity}
\end{figure*}

\begin{figure*}
    \centering
    \includegraphics{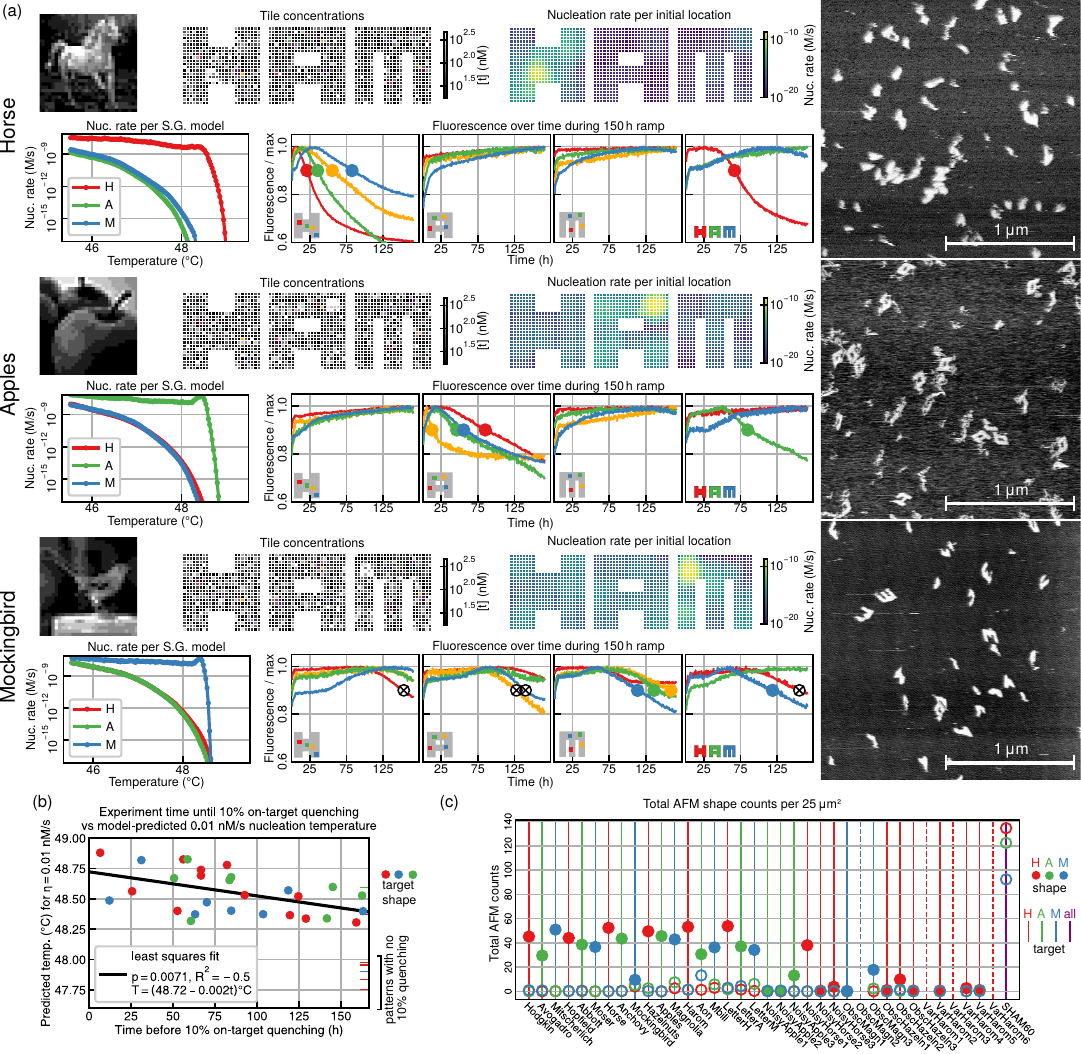}
    \caption{\textbf{Classification of images viewed as concentration patterns.}  36 different concentration patterns, derived from a mapping of 36 grayscale images, were run using a ramp between 48\,\degC to 45\,\degC over approximately 150 hours.
    \textbf{a,} Three pattern examples, with source image, concentration pattern, nucleation model nucleation rate starting from particular tiles, nucleation model nucleation rates, fluorescence results, and AFM images. \textbf{b,} Across all patterns there was some correlation between the on-target nucleation temperature predicted by the nucleation model and on-target shape quenching time. \textbf{c,} Total AFM shape counts for each sample.  Information for all patterns is available in the \sipatterndata.} 
    \label{extfig:patterns}
\end{figure*}

\begin{figure*}
    \centering
    \includegraphics{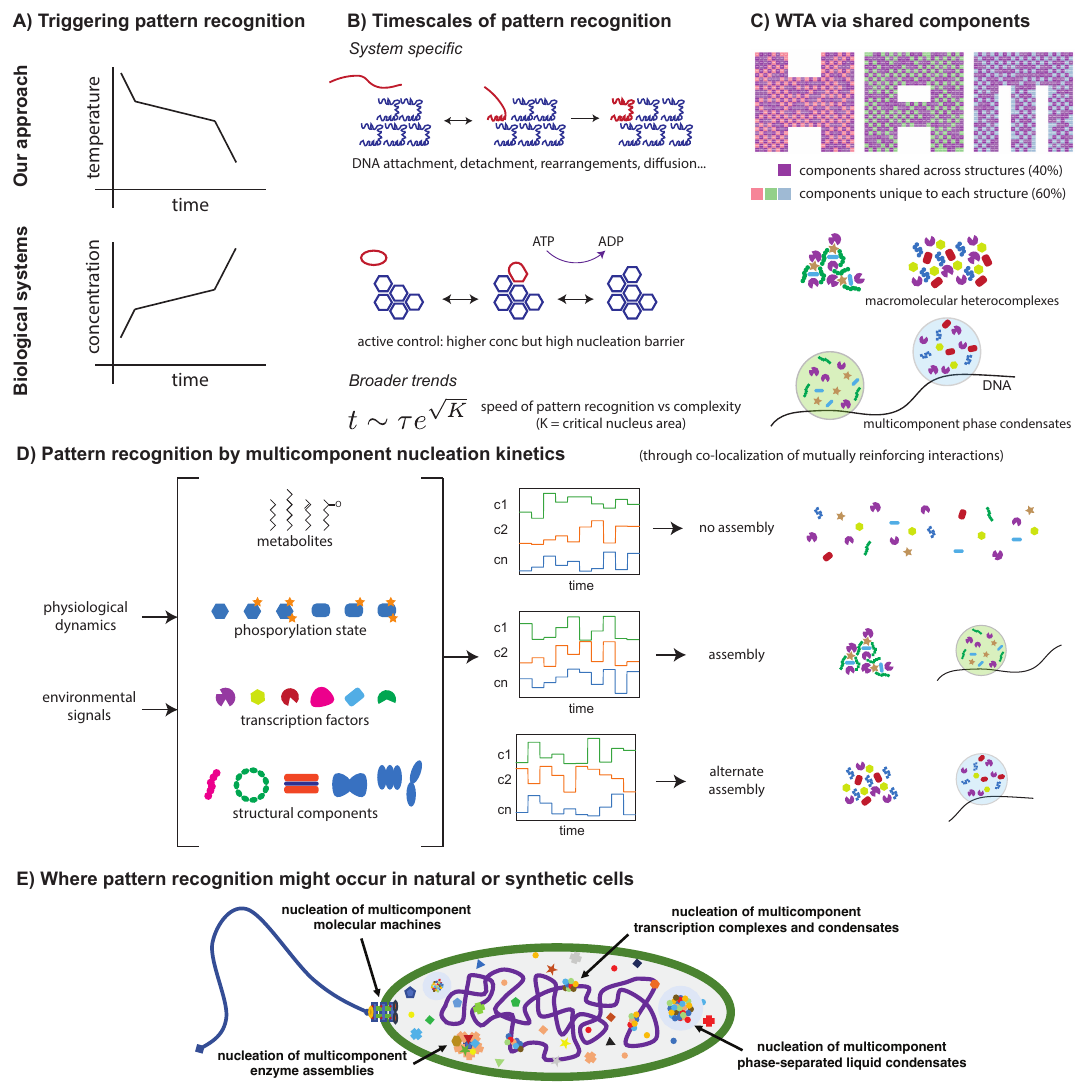}
    \caption{\textbf{ Parallels and differences between pattern recognition in our synthetic approach and in potential biological systems.}
    While we studied pattern recognition using a specific set of molecules (DNA) in an in vitro system, the concepts behind our work have potential relevance to biological systems built out of molecules of different nature and in different conditions. \textbf{a,}  Pattern recognition was triggered in our system by lowering temperature over time which drives the self-assembly process forward. Instead, in the cell, concentrations of molecular components can rise over time (e.g., through gene expression), leading to nucleation and self-assembly and thus pattern recognition.  \textbf{b,}  Timescale of pattern recognition is controlled by system-specific aspects and also general trends. System specific: DNA-specific processes such as tile attachment, detachment and restructuring set the timescale of nucleation and growth. Living systems can use active mechanisms to control nucleation timescales in addition to the concentration effects described here.  Broader trends: our theoretical work supports a general relationship between the speed of pattern recognition (e.g., by working at a lower temperature), the size of critical nuclei and thus the complexity of pattern recognition.  \textbf{c,}  The winner-take-all effect in our work enhanced selectivity by exploiting the depletion of shared components. Biomolecular systems, such as macromolecular complexes \autocite{sartori2020lessons} and multicomponent phase condensates \autocite{Hnisz2017-kg} are thought to share components as well, potentially enabling a winner-take-all effect in cells.  \textbf{d,}  In the biological context, the inputs could represent physiological or environmental signals encoded in the relative concentrations of many species of molecules. Some patterns of enhanced concentration may not lead to self-assembly or phase condensation if those components with enhanced concentration are not colocalized on a structure or reinforce a nucleation pathway for a condensate; but an alternative pattern of high concentrations could lead to assembly of one of several assemblies or condensates. \textbf{e,} Such sensitivity of kinetic pathways to concentration patterns can be exploited for complex decision-making in numerous aspects of cellular physiology, or may provide compact and robust control mechanisms for cell-scale molecular robots.  See also \sireservoir for how a pixel-to-tile map could be physically incarnated.
    \label{extfig:biology}
    }
\end{figure*}

\end{refsegment}

\end{document}